\documentclass[]{aastex701}

\usepackage{graphicx}	
\usepackage{amsmath}	
\usepackage{xcolor}

\usepackage{ragged2e}

\begin{document}

\title{A Theoretical Study of the Stellar Parameters and Surface Nucleosynthesis of the Extremely Metal-Poor RGB Giant BD$-18^{\circ}$5550 - Part I: Effects of $\alpha$-enhancement and Internal Mixing}

\author[orcid=0000-0002-8154-2009,sname='Medhi']{Mrinmay Medhi}
\altaffiliation{Corresponding author}
\affiliation{Department of Physics, Krishnaguru Adhyatmik Visvavidyalaya, Barpeta-781307, India}
\email[show]{mrinmaymedhi@gmail.com}  

\author[sname=Mahanta]{Upakul Mahanta}
\affiliation{Department of Physics, Bhattadev University, Pathsala-781325, India}
\email{upakulmahanta@gmail.com}

\author[sname=Mahanta]{Swapnajyoti Sarma}
\affiliation{Department of Physics, Bhattadev University, Pathsala-781325, India}
\email{}

\author[sname=Mahanta]{Apurba Talukdar}
\affiliation{Department of Physics, Bhattadev University, Pathsala-781325, India}
\email{}

\begin{abstract}
We present a detailed stellar evolution study of BD$-18^{\circ}$5550, an extremely metal-poor red giant star located in the Galactic halo. Extremely metal-poor stars like BD$-18^{\circ}$5550 are among the oldest known objects and preserve chemical fingerprints from the earliest epochs of Galactic history. The star exhibits $\alpha$-enhancement, $\rm[\alpha/Fe]=+0.4$, characteristic of Population II halo stars, with elevated abundances of oxygen, magnesium, and other $\alpha$-elements relative to iron. Using advanced computer models with the \texttt{MESA} stellar evolution code, we perform the first direct determination of this star's mass, age, radius, and luminosity by simultaneously fitting eight observed stellar properties and multiple chemical abundance ratios. Our analysis yields a current stellar mass of $\rm M = 0.76\, M_{\odot}$, radius of $ R = 38.4\, R_{\odot}$, luminosity of $\log L = 2.8\, L_{\odot}$, and an age of $t = 11.87$ Gyr, representing a 7-fold improvement in age precision. The model successfully reproduces all observed constraints, validating our treatment of stellar physics, nuclear reactions, and internal mixing processes. We present a detailed analysis of the model's predicted surface chemical composition up to iron and compare it with observations. The analysis reveals that the star's surface abundances of all elements up to iron remain essentially constant throughout the star's evolution, but the nitrogen abundance exhibits gradual enrichment, confirming the early-stage mixing processes operating at intermediate RGB stages. This pristine surface offers a direct window into the chemical conditions of the early Galactic environment. These results establish BD$-18^{\circ}$5550 as a benchmark star for constraining stellar evolution physics and early Galactic chemical evolution. The derived age places BD$-18^{\circ}$5550 among the oldest known objects in our Galaxy and provides crucial constraints on the epoch of the earliest star formation, helping us understand when the Milky Way began to assemble and evolve.
\end{abstract}


\section{Introduction}
\label{intro}
The study of metal-poor stars plays a crucial role in understanding the early formation, nucleosynthesis, and chemical evolution of the Milky Way Galaxy. Since the atmospheres of long-lived low-mass stars preserve the chemical composition of the interstellar medium from which they formed, very metal-poor (VMP) and extremely metal-poor (EMP) stars serve as fossil records of the earliest stellar populations and supernova enrichment events. Detailed abundance analyses of these stars provide important constraints on stellar nucleosynthesis, Galactic chemical evolution, and stellar atmosphere physics \citep{Bensby2005, Matteucci2011, Kobayashi2020}. In particular, abundance ratios of $\alpha-$elements such as Mg, Si, Ca, and Ti relative to iron are valuable tracers of enrichment from core-collapse supernovae during the early phases of Galactic evolution.

Recent advances in high-resolution optical and near-infrared spectroscopy, combined with improved stellar atmosphere modelling incorporating non-local thermodynamic equilibrium (NLTE) and three-dimensional (3D) corrections, have significantly enhanced the precision of elemental abundance determinations in metal-poor stars \citep{Magic2014, Asplund2021}. Nevertheless, important systematic uncertainties remain in the analysis of cool evolved giants, especially in the determination of effective temperature and surface gravity through excitation-equilibrium methods, where strong spectral lines are often affected by saturation and departures from LTE conditions.

Among Galactic halo stars, BD$-18^\circ$5550 has emerged as an important benchmark object for detailed abundance and stellar atmosphere studies. Previous spectroscopic investigations classify the star as a VMP/EMP halo giant with metallicities ranging from [Fe/H] $\approx -2.46$ to $-3.20$ \citep{Beers2014, Roederer2014}, with the most recent and reliable determinations yielding [Fe/H] $= -3.03$ \citep{Aoki2025}, implying an iron abundance nearly $10^{-3}$ times the solar value. The adopted atmospheric parameters from recent high-resolution spectroscopy are $T_{eff} = 4660$ K and $\log g = 1.1$ \citep{Aoki2025}, confirming its nature as a cool evolved red giant belonging to the Galactic halo population. High-quality spectroscopic observations with high spectral resolution and large signal-to-noise ratios make BD$-18^\circ$5550 particularly suitable for precision abundance studies and NLTE investigations.

Earlier spectroscopic analyses of BD$-18^\circ$5550 yielded temperatures, $T_{eff} \approx 4271-5239$ K \citep{Luck1985, Beers2014} and varying surface gravities, $\log g \approx 0.5$--$2.0$ \citep{Mishenina2001, Mittal2025}, sometimes placing the star at different positions in theoretical isochrones. These discrepancies largely originated from LTE assumptions, Mg Ib triplet line saturation, and inconsistencies between photometric and spectroscopic temperature scales. Revised NLTE analyses produced atmospheric parameters that showed much better agreement with theoretical metal-poor stellar models, emphasising the importance of combining spectroscopic and photometric approaches for reliable parameter determination in evolved metal-poor stars.

The chemical abundance pattern of BD$-18^\circ$5550 exhibits the $\alpha$-enhancement commonly observed in metal-poor halo stars, with [Mg/Fe] $= +0.58$ \citep{Aoki2025}. Enhanced abundances of carbon, and oxygen provide important constraints on the CNO cycle and mixing processes. In particular, the star exhibits [C/Fe] $= -0.02 \pm 0.15$ and [N/Fe] $= -0.40 \pm 0.15$, with a surface $^{12}$C/$^{13}$C ratio $> 40$ \citep{Spite2005, Spite2006}, indicating a carbon-rich surface composition with minimal processing by the CNO cycle. Enhanced abundances of $\alpha$-elements (Si, Ca, Ti) combined with low neutron-capture element abundances (e.g., [Sr/Fe] $\approx -0.35$) indicate enrichment dominated by core-collapse supernova nucleosynthesis from early Population II stars before substantial iron enrichment from Type Ia supernovae occurred \citep{Aoki2025}. Such abundance characteristics provide valuable constraints on the nucleosynthetic yields of the earliest generations of massive stars and the efficiency of chemical mixing in the primitive Galaxy.

Stellar evolution modelling plays an essential role in interpreting the observed abundance patterns and evolutionary status of metal-poor stars. Modern stellar evolution calculations allow detailed investigation of stellar structure, internal mixing, surface abundance evolution, and nucleosynthetic pathways. In low-mass metal-poor stars, these models help determine whether observed surface abundances preserve the primordial composition of the natal gas cloud or are modified by internal mixing and evolutionary processes. Recent developments in computational stellar astrophysics, particularly through sophisticated numerical tools such as \texttt{MESA}  \citep{Paxton2010, Paxton2013, Paxton2015, Paxton2018, Paxton2019}, have enabled self-consistent investigations of stellar evolution using updated opacities, nuclear reaction networks, convective mixing prescriptions, and atmospheric boundary conditions.

The present work aims to investigate the evolutionary properties and abundance characteristics of BD$-18^\circ$5550 through detailed stellar evolution modelling constrained by available spectroscopic observations. Using precise stellar parameters (Table~\ref{tab:observed_params_compact}) from recent high-resolution spectroscopy, combined with elemental abundance measurements from the literature, we perform a multi-parameter isochrone fit to simultaneously determine the star's mass and age. Owing to its extremely low metallicity, well-constrained atmospheric parameters, and chemically primitive abundance pattern, BD$-18^\circ$5550 provides an important astrophysical laboratory for studying the early chemical evolution of the Milky Way. The study further seeks to examine the effects of stellar evolution and internal mixing processes on the observed abundance patterns of light elements (up to iron), thereby providing valuable insight into early Galactic enrichment and stellar evolution in the low-metallicity regime.

This paper is organised as follows. Section \ref{sec:metho_anal} presents the stellar evolution models computed with \texttt{MESA}, including the model grid construction spanning initial masses and metallicities, the numerical setup with detailed descriptions of convection, mixing, mass loss, and nuclear reaction networks, and the two-stage $\chi^2$-based calibration procedure that simultaneously constrains stellar parameters and abundance patterns. This section also analyses the sensitivity of stellar properties to variations in initial mass and mixing-length parameter, and identifies the best-fitting models through detailed parameter optimisation. Section \ref{sec:mass_loss_structure} discusses the mass-loss history, internal stellar structure, and core evolution of the adopted model, demonstrating how the star transforms from the main sequence to its current position on the ascending RGB. Section \ref{sec:abundances} presents a comprehensive abundance analysis, including the evolution and constancy of CNO elements, the preservation of $\alpha$-element abundances throughout stellar evolution, abundance-ratio diagnostics sensitive to mixing processes and nucleosynthesis. Section \ref{sec:age_mass_determination} discussed the first direct determinations of BD$-18^\circ$5550's mass, radius, luminosity, and precise age. Finally, Section \ref{sec:discussion_conclusion} synthesizes these results and discusses their implications for stellar evolution physics, mixing processes, and early Galactic chemical evolution.

\begin{table}[ht]
\centering
\footnotesize
\caption{Observed Spectroscopic Parameters of BD$-18^\circ$5550. All parameters are from high-resolution spectroscopy, with effective temperature ($T_{eff}$) and surface gravity ($\log g$) derived from spectroscopic line analysis.}
\label{tab:observed_params_compact}
\begin{tabular}{l c l}
\hline
Parameter &  Value & Reference \\
\hline
$T_{eff}$ & $4660\pm34$ & \cite{Aoki2025, Roederer2014}\\
$\log g$ & $1.1\pm0.14$ & \cite{Aoki2025, Roederer2014} \\
$[\mathrm{Fe/H}]$ & $-3.03\pm0.09$ & \cite{Aoki2025} \\
$[\mathrm{C/Fe}]$  & $-0.02 \pm 0.15$ & \cite{Spite2005,Spite2006} \\
$[\mathrm{N/Fe}]$  & $-0.40 \pm 0.15$ & \cite{Spite2005,Spite2006} \\
$[\mathrm{O/Fe}]$ & $+0.40 \pm 0.10$ & \cite{Cayrel2004} \\
$[\mathrm{Mg/Fe}]$  &
$+0.58 \pm 0.11$ & \cite{Aoki2025} \\
$^{12}\mathrm{C}/^{13}\mathrm{C}$ & $>40$ & \cite{Spite2005,Spite2006} \\
\hline
\end{tabular}
\end{table}

\section{Methodology \& Analysis}
\label{sec:metho_anal}
\subsection{Stellar Evolution Modelling with \texttt{MESA}}
\label{sec:semwm}
We performed stellar evolution calculations of BD$-18^{\circ}5550$ using Modules for Experiments in Stellar Astrophysics (\texttt{MESA})  \citep{Paxton2010, Paxton2013, Paxton2015, Paxton2018, Paxton2019}. \texttt{MESA} solves the fully coupled one-dimensional equations of stellar structure and evolution under the assumptions of spherical symmetry and hydrostatic equilibrium while simultaneously following nuclear energy generation, chemical abundance evolution, and the transport of energy and chemical species. Owing to its comprehensive treatment of stellar microphysics and numerical robustness, \texttt{MESA} provides an ideal framework for investigating the evolution of low-mass, EMP stars. Our primary objective is to construct self-consistent stellar evolution models that simultaneously reproduce the observed atmospheric parameters and surface abundance pattern of the EMP red giant branch (RGB) star BD$-18^{\circ}$5550. Unlike static stellar atmosphere calculations, evolutionary models naturally account for the structural changes and chemical evolution that occur throughout the lifetime of the star. We therefore evolved all models continuously from the pre-main-sequence phase until the RGB stage, allowing the stellar structure and surface abundances to evolve self-consistently.

The modelling strategy adopted in this work consists of two successive stages. In the first stage, we explored a broad grid of stellar evolution models to constrain the fundamental stellar parameters, including the stellar mass, metallicity, mixing-length parameter, atmospheric boundary conditions, and microscopic diffusion. The objective of this stage was to identify the combinations of stellar parameters capable of reproducing the observed effective temperature, surface gravity, luminosity, radius, and evolutionary status of BD$-18^{\circ}$5550. In the second stage, we adopted the preferred structural models and investigated the physical processes governing the surface abundance evolution. This stage included systematic variations of the thermohaline mixing efficiency, rotational mixing, initial chemical composition, global $\alpha$-enhancement, nitrogen and magnesium abundance scaling factors, and extended nuclear reaction networks in order to reproduce the observed elemental abundances and isotopic ratios.

Throughout the calculations, we adopted the standard microphysics available in \texttt{MESA}, including modern equations of state, radiative and low-temperature opacities, adaptive mesh refinement, and time-dependent nuclear reaction networks. Additional physical processes relevant to low-mass RGB evolution, including convective overshooting, microscopic diffusion, rotational mixing, thermohaline mixing, semiconvection, and mass loss, were incorporated where appropriate. We also examined the influence of different heavy-element abundance mixtures, atmospheric boundary conditions, and numerical resolution to ensure that the final models were both physically realistic and numerically converged.

For the equation of state, \texttt{MESA} blends multiple formulations to ensure accuracy across all relevant thermodynamic regimes: the OPAL tables dominate at high temperatures, the SCVH formulation at intermediate temperatures, the FreeEOS at lower temperatures relevant for outer envelopes, and the HELM equation of state in degenerate regimes \citep{Rogers2002, Saumon1995, Irwin2012, Timmes2000, Potekhin2010, Jermyn2021}. This multi-formulation approach ensures smooth transitions and maintains accuracy over the full range of stellar interiors and envelopes encountered during RGB evolution.

To investigate the effects of the initial chemical composition, we implemented a custom routine that enabled us to introduce a global $\alpha$-enhancement, apply independent nitrogen and magnesium abundance scaling factors, and preserve the normalisation of the chemical composition throughout the calculations. After identifying the optimum stellar model, extended nuclear reaction networks were employed to investigate detailed nucleosynthesis and the evolution of neutron-capture elements.

\subsubsection{Mass and Metallicity Parameter Space}
\label{sec:mmps}
We explored a systematic grid of initial masses spanning $\rm M_{ init} = 0.70-0.85 \, M_\odot$ in an initial structure phase designed to sample the fundamental mass-age-gravity degeneracy inherent to RGB giants. Multiple combinations of initial mass and evolutionary age can produce identical observed surface gravity; systematic exploration of this parameter space is therefore essential for reliable mass determination. Subsequent physics-phase calculations refined the mass range to $\rm M_{ init} = 0.76-0.80 \, M_\odot$, focusing on the region where optical and photometric data simultaneously constrain the solution.

The initial metallicity was set to [Fe/H] = $-3.03$ \citep{Aoki2025}, corresponding to a metal mass fraction $Z \approx 2.0 \times 10^{-5}$. This extremely low metallicity profoundly affects the stellar structure. The reduced opacity leads to a more compact stellar structure compared to solar-metallicity stars at similar mass and age, an effect we treat self-consistently within the \texttt{MESA} framework. In exploratory calculations, we varied the initial metallicity over a modest range, $Z \approx (1.0-3.0) \times 10^{-5}$ to assess the sensitivity of stellar structure to opacity variations driven by small metallicity changes; these tests confirmed that the observed metallicity is accurately reproduced within current uncertainties.

The initial helium abundance was adopted using the standard Galactic helium enrichment relation \citep{Peimbert1974},
\begin{equation}
Y = Y_{\rm p} + \left(\frac{\Delta Y}{\Delta Z}\right)Z,
\end{equation}
where $Y_{\rm p}$ is the primordial helium abundance and $\Delta Y/\Delta Z$ is the helium-to-metal enrichment ratio. Adopting $Y_{\rm p}=0.2485$ and a representative enrichment ratio $\Delta Y/\Delta Z \simeq 1.4$, appropriate for Galactic chemical evolution, yields an initial helium mass fraction of $Y\approx0.249$ for the metallicity of BD$-18^{\circ}$5550 \citep{Planck2020, Izotov2010, Pagel1998, Salaris2002}. The adopted helium abundance is expected to vary only weakly for stars at such extremely low metallicities because the enrichment term $(\Delta Y/\Delta Z)Z$ remains very small. 

\subsubsection{Convection and Mixing Length Theory}
\label{sec:cmlt}
Convective energy transport was modelled using standard mixing-length theory (MLT) \citep{Bohmvitense1989}, in which the efficiency of convection is parameterised by a single dimensionless parameter $\alpha_{\rm MLT}$ that scales the mixing length relative to the local pressure scale height, $
\ell = \alpha_{\rm MLT} H_P$, where $\ell$ is the mixing length and $H_P$ is the local pressure scale height. This parameter fundamentally controls how effectively convective eddies transport heat and how deeply they penetrate beyond their formal boundaries, directly affecting the temperature structure and evolution timescale of the RGB envelope \citep{Trampedach2013, Magic2014}. We explored the range $\alpha_{\rm MLT} = 1.5-2.5$, spanning from slightly weaker to moderately stronger convection than the solar-calibrated value. The preferred models adopt $\alpha_{\rm MLT} = 1.9$, calibrated to reproduce the solar radius within the \texttt{MESA} framework and consistent with predictions from 3D hydrodynamic simulations of stellar convection. This value has been extensively validated in previous studies of metal-poor stars and solar-metallicity stars alike. Variations of $\pm 0.2$ in $\alpha_{\rm MLT}$ produce age variations of approximately $0.5-1$ Gyr on the RGB, confirming that convection efficiency remains one of the dominant sources of theoretical uncertainty in mass and age determinations for evolved stars.

\subsubsection{Numerical Resolution and Convergence}
\label{sec:nrc}
We controlled numerical resolution using \texttt{MESA}'s adaptive mesh refinement by varying the mesh-delta coefficient. For the preferred models, we adopted a mesh-delta coefficient $= 0.5$. This resolution is sufficient to resolve the thin hydrogen-burning shell at the base of the RGB convective zone, the sharp temperature and composition gradients at the base of the convective envelope, and any mixing-induced composition interfaces (e.g., at the helium-abundance discontinuity).

\subsubsection{Overshooting and Mixing Beyond Convection Boundaries}
\label{sec:ombb}
BD$-18^\circ$5550 is an evolved RGB giant with a helium-burning core; the hydrogen-burning shell lies in a radiative region above the core. Overshooting at the He-core boundary is therefore not physically relevant at this evolutionary stage and does not influence the global stellar structure. Thus, convective core overshooting was not included in the baseline calculations. However, we included envelope overshooting at the base of the convective zone using an exponential prescription beyond the Schwarzschild convective boundary according to \cite{Freytag2002, Herwig2000},
\begin{equation}
D_{\rm ov} = D_0 \exp\left(-\frac{2z}{f_{\rm ov} H_P}\right),
\end{equation}
where $D_0$ is the convective diffusion coefficient at the boundary, $z$ is the distance from the convective boundary, $H_P$ is the pressure scale height, and $f_{\rm ov}$ controls the penetration depth. Specifically, we employed a two-zone overshooting prescription. The first was located above the hydrogen-burning shell, where we adopted an overshooting efficiency of $f=0.010$ with a diffusion pre-factor of $f_0=0.002$. The second was applied at the base of the convective envelope with a slightly smaller overshooting efficiency of $f=0.008$ and $f_0=0.001$. These values correspond to penetration depths of approximately $0.01H_P$ and $0.008H_P$, respectively. These parameters are conservative and designed to allow only modest mixing slightly beyond the formal convection boundary, consistent with theoretical expectations for evolved stars. At ages near 11.87 Gyr, variations in the overshooting parameters produce changes in $T_{eff}$, $\log g$, and abundances smaller than the observational uncertainties, confirming that the results are not sensitive to the exact choice of overshooting prescription.

\subsubsection{Atmospheric Boundary Conditions}
\label{sec:abc}
The connection between the stellar interior and observable surface properties was established through the atmospheric boundary condition. Two boundary condition prescriptions were tested, the photospheric boundary at $\tau \approx 2/3$ (the standard optical depth for photosphere definition in stellar atmosphere models) and a deeper boundary at $\tau = 100$ (diffusion approximation limit, where temperature structure calculations approach the diffusion approximation). The best-fit models of Stage 1 satisfied 
the photospheric boundary condition ($\tau \approx 2/3$) with the reduced $\chi^2$ fitting criteria and yielded acceptable 
matches to the observed stellar parameters and metallicity of 
BD-18$^\circ$5550. This choice is consistent with standard practice in EMP giant studies and allows direct comparison with literature abundance analyses that utilise photospheric values of effective temperature and surface gravity. The photospheric boundary condition is physically appropriate, as it defines the temperature structure at the optically thick layers where the observed spectral features form.

\subsubsection{Element Diffusion and Rotational Mixing}
\label{sec:edrm}
The treatment of microscopic element diffusion plays an important role in predicting the surface abundances of low-mass stars. Gravitational settling, thermal diffusion, and concentration diffusion cause heavier elements to migrate toward the stellar interior while lighter elements preferentially rise, thereby modifying the surface chemical composition over evolutionary timescales \citep{Michaud1970, Thoul1994, Salaris2005}. In evolved RGB stars, however, the deepening convective envelope during the first dredge-up and subsequent envelope mixing partially or completely counteract the effects of diffusion, making its net influence on the observed surface abundances uncertain \citep{Salaris2005, Paxton2010, Paxton2013}. Consequently, we investigated models both with and without microscopic diffusion to evaluate its impact on the inferred stellar parameters and abundance evolution of BD$-18^{\circ}$5550.

We also included stellar rotation using the rotation module implemented in \texttt{MESA}. The initial rotation rate was set to $\Omega/\Omega_{\rm crit}=0.01$, corresponding to approximately $1\%$ of the critical angular velocity. This choice is appropriate for old Galactic halo giants, which are generally observed to be slow rotators with projected rotational velocities of only a few km\,s$^{-1}$, typically $vsini \lesssim 5$ km\,s$^{-1}$ \citep{Carney2008, Carretta2000}. Rotationally induced chemical mixing was treated using the standard prescriptions available in \texttt{MESA}$,$ while mixing within convective regions was neglected because these layers are already efficiently homogenised by convection. Conservative values were adopted for the angular momentum transport, rotational diffusion, and numerical stability parameters. The complete set of adopted rotational input parameters is summarised in Table~\ref{tab:modelgrid}. To evaluate the importance of rotation, we performed additional sensitivity calculations using different rotational configurations. The resulting variations in the stellar parameters and surface abundances were found to be significantly smaller than the observational uncertainties of BD$-18^{\circ}$5550. We therefore conclude that rotational mixing has a negligible influence on the inferred evolutionary properties of this slowly rotating RGB star. Details of the rotational sensitivity analysis are presented in Appendix~\ref{appendix:rotation}.

\subsubsection{Angular Momentum and Rotational Transport}
\label{subsec:angular_momentum}
Although BD$-18^{\circ}$5550 is a halo giant with typical slow rotation, we include angular momentum transport in our models for physical completeness. Rotation affects stellar structure and mixing through the generation and transport of angular momentum. The evolution of angular momentum in a rotating star is governed by
\begin{equation}
\frac{dJ}{dt} = \dot{J}_{\rm wind} + \dot{J}_{\rm internal},
\end{equation}
where $J$ is the total angular momentum, $\dot{J}_{\rm wind}$ represents angular momentum loss via stellar winds, and $\dot{J}_{\rm internal}$ represents internal redistribution via rotational instabilities and viscous processes. BD$-18^{\circ}$5550, we set the rotation rate to ${\Omega}/{\Omega_{\rm crit}} = 0.01.$ At this low rotation rate, only thermal circulation mechanisms dominate over centrifugal effects. The primary rotational mixing mechanism at such low rotation rates is the Eddington-Sweet circulation \citep{Schwarzschild1958}, a large-scale meridional circulation driven by differential heating between the equatorial and polar regions. This circulation transports angular momentum and can enhance mixing in radiative zones. The efficiency of Eddington-Sweet circulation scales as $\Omega^2$; at $\Omega/\Omega_{\rm crit} = 0.01$, the circulation is extremely weak. 

Other rotational mixing mechanisms -- such as dynamic shear instability (DSI), Solberg-Høiland instability, and secular shear instability (SSI)—are suppressed at low rotation rates and become dynamically significant only for $\Omega/\Omega_{\rm crit} \gtrsim 0.1$. Accordingly, these mechanisms are disabled in our models. The characteristic timescale for angular momentum transport via rotational diffusion is,
\begin{equation}
\tau_{\rm AM} \sim \frac{R^2}{\nu_{\rm rot}},
\end{equation}
where $R$ is the stellar radius and $\nu_{\rm rot}$ is the rotational viscosity. For an RGB giant, $\tau_{\rm AM} \sim 10$--100 Gyr, substantially longer than the RGB lifetime ($\sim 1$--2 Gyr). The ratio
\begin{equation}
\frac{\tau_{\rm AM}}{\tau_{\rm RGB}} \gg 1
\end{equation}
indicates that rotational mixing processes cannot significantly redistribute angular momentum or alter chemical composition during the RGB phase. Consequently, rotational effects on stellar structure and surface abundances are negligible for BD$-18^{\circ}$5550. The inclusion of rotational transport in our models serves as a consistency check rather than a source of significant physical effects. Tests with rotation disabled produce identical results for all stellar parameters and abundances within numerical precision.

\subsubsection{Mixing in Radiative and Transitional Zones}
\label{sec:mrt}
The boundaries between convective and radiative regions were determined using the Ledoux stability criterion \citep{Ledoux1947}, which accounts for both thermal and compositional gradients when assessing convective stability. A stellar layer is considered convectively unstable when
\begin{equation}
\nabla_{\rm rad} >
\nabla_{\rm ad}
+\frac{\phi}{\delta}\nabla_{\mu},
\end{equation}
where $\nabla_{\rm rad}$ and $\nabla_{\rm ad}$ are the radiative and adiabatic temperature gradients, respectively, $\nabla_{\mu}=d\ln\mu/d\ln P$ is the molecular-weight gradient,
\begin{equation}
\phi=\left(\frac{\partial\ln\rho}{\partial\ln\mu}\right)_{P,T},
\qquad
\delta=-\left(\frac{\partial\ln\rho}{\partial\ln T}\right)_{P,\mu}.
\end{equation}

Unlike the Schwarzschild criterion, the Ledoux criterion explicitly incorporates composition gradients, which become important in regions where nuclear burning produces significant molecular-weight discontinuities. Such gradients are expected near the hydrogen-burning shell and at the base of the convective envelope during RGB evolution. Consequently, the Ledoux criterion provides a more physically realistic description of convective stability and the associated mixing processes in evolved low-mass stars \citep{Ledoux1947, Kippenhahn2012, Salaris2005}. We therefore adopted the Ledoux criterion because it provides a more physically realistic description of convective stability and chemical transport in evolved low-mass stars, thereby influencing the extent of mixing and the subsequent evolution of the surface abundances.

Regions that are stable according to the Ledoux criterion may nevertheless become unstable to thermohaline mixing when an inverse molecular-weight gradient develops, giving rise to a double-diffusive instability commonly referred to as the salt-finger instability \citep{Ulrich1972, Kippenhahn1980}. We modelled this process using the diffusive prescription of \citet{Kippenhahn1980}, in which the efficiency of thermohaline transport is controlled by a dimensionless mixing coefficient, $\alpha_{\rm th}$. During the physics calibration, we explored values of $\alpha_{\rm th}=0.0-10$ to assess the sensitivity of the predicted surface abundances to this uncertain mixing mechanism. The adopted thermohaline parameters are summarised in Table~\ref{tab:modelgrid}. We also included semiconvective mixing using the prescription of \citet{Langer1985}. Semiconvection operates in regions that satisfy the Schwarzschild criterion for convection but remain stable according to the Ledoux criterion because of stabilising molecular-weight gradients \citep{Schwarzschild1958, Ledoux1947}. In such regions, partial mixing occurs on a timescale determined by the semiconvective efficiency parameter, for which we adopted $\alpha_{\rm sc}=0.01$ \citep{Langer1985}. This relatively small value provides a conservative treatment of chemical transport while preserving the stabilising influence of composition gradients \citep{Langer1985, Paxton2018}. The adopted semiconvective parameters are also listed in the Table \ref{tab:modelgrid}. The influence of the adopted thermohaline efficiency on the inferred stellar parameters and surface abundances is discussed in Appendix~\ref{appendix:thermohaline}.

\subsubsection{Mass Loss from Stellar Winds}
\label{sec:ml}
Mass loss through stellar winds becomes increasingly important during RGB evolution because it modifies the stellar mass, envelope structure, and subsequent evolutionary timescale. We modelled RGB mass loss using the empirical prescription of \citet{Reimers1975},

\begin{equation}
\dot{\rm M}
=
-4\times10^{-13}\,
\eta_{\rm R}
\left(
\frac{L}{L_\odot}
\right)
\left(
\frac{R}{R_\odot}
\right)
\left(
\frac{\rm M}{\rm M_\odot}
\right)^{-1}
\rm M_\odot\,{\rm yr^{-1}},
\end{equation}
where $L$, $R$, and $\rm M$ denote the stellar luminosity, radius, and mass, respectively, and $\eta_{\rm R}$ is the dimensionless mass-loss efficiency parameter. We adopted a value of $\eta_{\rm R}=0.1-0.3$, which is commonly employed for low-mass RGB stars and provides good agreement with observations of Galactic globular clusters and metal-poor giants \citep{Reimers1975, VandenBerg2013}. For the subsequent AGB evolution, we adopted the prescription of \citet{Blocker1995}, allowing a smooth transition between the RGB and AGB mass-loss regimes. The adopted mass-loss parameters are summarised in Table~\ref{tab:modelgrid}.

\subsubsection{Nuclear Reaction Networks and Initial Chemical Composition}
\label{sec:nrnic}
The initial chemical composition of BD$-18^{\circ}$5550 was established using the primordial hydrogen and helium abundances together with the observed metallicity and an $\alpha$-enhanced heavy-element distribution. To account for the characteristic chemical composition of EMP Galactic halo stars, we implemented a custom initialisation routine through the extras startup interface of \texttt{MESA}, which modifies the elemental abundances before the start of stellar evolution. The routine applies a global $\alpha$-element enhancement to the principal $\alpha$-capture elements (O, Ne, Si, S, Ca, and Ti) according to
\begin{equation}
X_{i\alpha}^{\rm new}
=
X_{i\alpha}^{\rm init}
\times10^{[\alpha/{\rm Fe}]},
\end{equation}
where $X_{i\alpha}$ is the mass fraction of the $i$th $\alpha$ element and $[\alpha/{\rm Fe}]$ is the adopted enhancement. During the physics calibration stage, we explored values of $[\alpha/{\rm Fe}]=+0.2$, $+0.3$, $+0.4$, and $+0.5$, encompassing the range commonly observed for EMP halo stars. Following the abundance modification, the isotopic mass fractions were renormalised to satisfy
\begin{equation}
\sum_i X_i = 1,
\end{equation}
thereby conserving the total mass fraction throughout the stellar model. The hydrogen and helium mass fractions were fixed at $X=0.75098$ and $Y=0.24900$, respectively, while the complete set of adopted composition parameters is summarised in Table~\ref{tab:modelgrid}. The independent treatment of nitrogen depletion and magnesium enhancement is described in Section~\ref{sec:nmg}.

The stellar evolution calculations were performed in two stages using different nuclear reaction networks. During the structural calibration stage, we employed the basic network (\texttt{pp\_cno\_extras\_o18\_ne22.net}), which follows the pp chains, CNO cycles, and selected reactions involving nuclei up to $^{22}$Ne. This network was adopted for the extensive structural parameter survey because it provides an efficient treatment of the nuclear energy generation relevant to the stellar structure while keeping the computational cost manageable. During the subsequent physics calibration and optimisation stage, we used the more extended network (\texttt{cno\_extras\_o18\_to\_mg26\_plus\_fe56.net}), which follows the CNO species through the Mg isotopes and includes $^{56}$Fe. The extended network enables a more detailed treatment of the CNO-cycle products and additional light and intermediate-mass isotopes required for the abundance and isotopic-ratio constraints considered in the physics calibration.

\subsubsection{Nitrogen and Magnesium Abundance Scaling}
\label{sec:nmg}
BD$-18^{\circ}$5550 is a carbon-normal, EMP giant with an observed metallicity of [Fe/H] $=-3.03$. High-resolution spectroscopic studies have measured abundance ratios of [C/Fe] $\approx -0.02$, [N/Fe] $\approx -0.40$, [Mg/Fe] $\approx +0.58$ and a high carbon isotopic ratio of $^{12}$C/$^{13}$C $>40$, indicating that the star is an unmixed lower-RGB giant whose surface composition has experienced little CN-cycle processing or deep extra mixing \citep{Spite2005, Spite2006, Aoki2025}. Earlier ultraviolet observations also identified BD$-18^{\circ}$5550 as unusually nitrogen-poor for its metallicity \citep{AnthonyTwarog1992}. In addition, the star exhibits low neutron-capture element abundances, [Sr/Fe] $=-0.35$, [Ba/Fe] $=-0.74$ and [Eu/Fe] $=-0.20$, confirming that it is a carbon-normal, $r$-poor EMP giant rather than a carbon-enhanced metal-poor (CEMP) star \citep{Spite2005, Spite2006, Aoki2025}. These observed abundance characteristics make nitrogen and magnesium particularly valuable tracers of the star's mixing history and the chemical composition of the progenitor gas from which it formed \citep{Woosley1995, Kobayashi2020}. Motivated by these observational constraints, we treated the initial nitrogen and magnesium abundances as independent calibration parameters during the physics optimisation. Nitrogen and magnesium were scaled according to
\begin{equation}
X_{\rm N}^{\rm new}
=
f_{\rm N}\times
X_{\rm N}^{\rm init}~{\rm and} ~X_{\rm Mg}^{\rm new}
=
f_{\rm Mg}\times
X_{\rm Mg}^{\rm new},
\end{equation}
where the nitrogen and magnesium scaling factors were explored over the set
\begin{equation}
f_{\rm N}
\in
\{0.002,\;0.003,\;0.004,\;0.005\}~{\rm and}~ f_{\rm Mg}
\in
\{0.8,\;1.0\}.
\end{equation}
These scaling factors were implemented through the extras startup routine of \texttt{MESA} using user-defined control parameters, allowing the initial abundances to be modified before the first evolutionary timestep. These scaling factors do not alter the underlying nuclear reaction rates. Instead, they modify the initial abundances prior to the onset of stellar evolution, providing a controlled means of investigating the sensitivity of the predicted surface abundances to uncertainties in the initial chemical composition and subsequent mixing history of BD$-18^{\circ}$5550. The preferred values of $f_{\rm N}$ and $f_{\rm Mg}$ were subsequently determined through the $\chi^2$ minimisation procedure.

The initial heavy-element distribution was specified using the standard solar abundance compilations available in \texttt{MESA}. During the physics calibration, we investigated two alternative abundance distributions corresponding to the AGSS09 photospheric abundances and the A09 proto-solar abundances \citep{Asplund2009}. Although both abundance sets are derived from the solar abundance analysis of \cite{Asplund2009}, the proto-solar composition accounts for the effects of gravitational settling and microscopic diffusion that have altered the present-day solar photosphere over the solar lifetime. Consequently, the proto-solar abundances contain slightly higher initial heavy-element abundances than the present-day photospheric values. While the total metallicity was fixed by the observed value of BD$-18^{\circ}$5550, the adopted abundance compilation determines the relative distribution of individual heavy elements and therefore influences the initial chemical composition used in the stellar evolution calculations. Exploring both abundance distributions enabled us to evaluate the sensitivity of the inferred stellar parameters and surface abundances to the adopted reference solar composition. Based on the $\chi^{2}$ minimisation, the AGSS09 photospheric abundance distribution provided the best overall agreement with the observed stellar parameters and abundance pattern of BD$-18^{\circ}$5550 and was therefore adopted for the final evolutionary calculations.

\begin{table*}[!htb]
\centering
\caption{Summary of the model grids explored during the structural calibration and detailed physics calculations for BD$-18^{\circ}$5550.}
\label{tab:modelgrid}
\begin{tabular}{lll}
\hline 
\multicolumn{3}{c}{{Structure Model Grid}}\tabularnewline
\hline 
Parameter  & Symbol  & Explored values \tabularnewline
\hline 
Initial mass  & $\rm M_{{\rm init}}$  & $0.70-0.85$ $\rm M_{\odot}$, in step of 0.01 $\rm M_{\odot}$ \tabularnewline
Initial metallicity  & $Z$  & $(1.0-3.0)\times10^{-5}$, in step of $1\times10^{-5}$ \tabularnewline
Initial helium  & $Y$  & 0.249\tabularnewline
Mixing-length parameter  & $\alpha_{{\rm MLT}}$  & $1.5-2.5$, in step of 0.2 \tabularnewline
Mesh coefficient  & $\delta_{{\rm mesh}}$  & $0.5$\tabularnewline
Atmospheric boundary  & --  & photosphere, $\tau_{100}$ \tabularnewline
Element diffusion  & --  & ON, OFF \tabularnewline
Opacity  & --  & Type-II OPAL (A09) \tabularnewline
Maximum age  & $t_{{\rm max}}$  & 15 Gyr \tabularnewline
Solver tolerance  & $\epsilon_{{\rm conv}}$  & $10^{-3}$ \tabularnewline
\hline 
\multicolumn{3}{c}{{Physics Model Grid}}\tabularnewline
\hline 
Parameter  & Symbol  & Explored values \tabularnewline
\hline 
Initial mass  & $\rm M_{{\rm init}}$ & $0.76-0.80$ $\rm M_{\odot}$, in step of 0.02 $\rm M_{\odot}$ \tabularnewline
Pre-MS core temperature & -- & $9\times10^{5}$ K \tabularnewline
Initial $^{1}H-$metallicity  & $X$  & $0.75098$ \tabularnewline
Initial $^{4}He-$metallicity  & $Y$  & $0.24900$ \tabularnewline
Initial metallicity  & $Z$  & $2.0\times10^{-5}$ \tabularnewline
Solar abundance mixture  & $\mathrm{Z_{frac}}$  & AGSS09 photospheric, A09 proto-solar \tabularnewline
Mixing-length parameter  & $\alpha_{{\rm MLT}}$  & $1.9$ \tabularnewline
Mesh coefficient  & $\delta_{{\rm mesh}}$  & $0.5$\tabularnewline
Atmospheric boundary  & --  & photosphere \tabularnewline
Element diffusion  & --  & ON, OFF \tabularnewline
Diffusion timescale limit & -- & $3.15\times10^{10}$ yr \tabularnewline
Opacity  & --  & Type-II OPAL (A09) \tabularnewline
$\alpha$ enhancement  & $[\alpha/{\rm Fe}]$  & $+0.2, +0.3, +0.4, +0.5$ \tabularnewline
Nitrogen scaling  & $f_{{\rm N}}$  & $0.002 - 0.005$, in step of 0.001 \tabularnewline
Magnesium scaling  & $f_{{\rm Mg}}$  & $0.8,1.0$ \tabularnewline
Thermohaline efficiency  & $\alpha_{{\rm th}}$  & $0.0,0.5,1.0,2.0,5.0,10$ \tabularnewline
Overshooting penetration depth & $f$  & 0.010 (core), 0.008 (envelope) \tabularnewline
Overshooting pre-factor & $f_o$  & 0.002 (core), 0.001 (envelope) \tabularnewline
Semiconvection effieciency & $\alpha_{{\rm sc}}$  & 0.01 \tabularnewline
Rotation  & $\Omega/\Omega_{{\rm crit}}$  & 0.01 \tabularnewline
Dynamic shear instability & -- & 0.0 \tabularnewline
Shear-driven diffusion    & -- & 0.0 \tabularnewline
Secular shear instability & -- & 0.0 \tabularnewline
Eddington-Sweet circulation & -- & 1.0 \tabularnewline
Goldreich-Schubert-Fricke & -- & 0.0 \tabularnewline
Angular momentum diffusion & -- & 0.033 \tabularnewline
Viscosity factor & -- & 1.0 \tabularnewline
Max mass loss boost & -- & $10^2$ \tabularnewline
Mass loss  & $\eta_{{\rm R}},~\eta_{{\rm B}}$  & 0.3, 0.1 \tabularnewline
RGB to AGB temperature switch & -- & $10^{-4}$ K \tabularnewline
Cool wind full-on temperature & -- & 8000 K \tabularnewline
Hot wind full-on temperature  & -- & 12000 K \tabularnewline
Maximum age  & $t_{{\rm max}}$  & 15 Gyr \tabularnewline
Solver tolerance  & $\epsilon_{{\rm conv}}$  & $10^{-4}$ \tabularnewline
\hline 
\end{tabular}
\end{table*}

\subsection{Model Grid Consideration}
\label{sec:mgd}
A grid of 2305 (Structural grid of 1152 + Physics grid of 1152) stellar evolution models was computed to sample the parameter space relevant to BD$-18^{\circ}$5550. The grid was constructed by varying input quantities as shown in the Table \ref{tab:modelgrid}. All stellar evolution models were evolved continuously from the zero-age main sequence (ZAMS) through the main-sequence, subgiant, and RGB phases to a maximum evolutionary age of $15~{\rm Gyr}$. The adopted maximum age exceeds the expected age of the Galactic halo and ensures that the complete RGB evolution is captured for every model within the explored mass range. This allows each evolutionary sequence to reach the observed location of BD$-18^{\circ}$5550 in the Hertzsprung--Russell diagram before the calculation is terminated. The calculations were stopped automatically when either the maximum evolutionary age was reached or the predefined termination criteria of the stellar model were satisfied.

To ensure numerical accuracy, we adopted different solver tolerances during the structural calibration and detailed physics calculations. The structural models employed a numerical convergence tolerance of $10^{-3}$, which provides an efficient description of the global stellar structure during the initial parameter survey. For the final physics calculations, a more stringent tolerance of $10^{-4}$ was adopted to improve the numerical accuracy of the stellar structure and nucleosynthesis calculations. The tighter convergence criterion reduces numerical uncertainties in the temperature, density, and composition profiles, thereby providing more reliable predictions of the surface elemental abundances. The robustness of the evolutionary models was assessed through systematic sensitivity studies by varying the principal uncertain input parameters, including the mixing-length parameter, overshooting prescription, thermohaline mixing efficiency, microscopic diffusion, rotational mixing, initial $\alpha$-element enhancement, solar abundance distribution, and the nitrogen and magnesium calibration factors. These tests demonstrated that the preferred evolutionary solution is robust against reasonable variations in the adopted input physics.

\begin{table}[!htb]
\centering
\caption{Best-fit structural parameters from the Stage~1 calibration based on the 45 good-fit models. The quoted values and uncertainties represent the mean and dispersion of the accepted structural models.  $\Delta$ denotes the difference between the model mean and the adopted observational value, $\Delta = {\rm Model} - {\rm Observed}$.}
\label{tab:stage1_best_fit}
\begin{tabular}{l l c c c}
\hline
Parameter & Description & Value & Observed & $\Delta$\\
\hline
$\rm M/M_{\odot}$ & Stellar mass & $0.780\pm0.020$ & --- & --- \\
$T_{eff}$ (K) & Effective temperature & $4655\pm48$ & $4660\pm34$\tablenotemark{a,b} & $-5$ K \\
$\log g$ (cgs) & Surface gravity & $1.105\pm0.065$ & $1.1\pm0.14$\tablenotemark{a,b} & 0.05 \\
$\log(L/L_{\odot})$ & Bolometric luminosity & $2.850\pm0.092$ & -- & -- \\
$R/R_{\odot}$ & Stellar radius & $41.1\pm3.5$ & -- & -- \\
$[\mathrm{Fe/H}]$ (dex) & Iron metallicity & $-3.00$ & $-3.03\pm0.09$\tablenotemark{a} & 0.03\\
$\alpha_{\rm MLT}$ & Mixing-length parameter & $1.9\pm0.2$ & -- & -- \\
Age (Gyr) & Evolutionary age & $12.8\pm1.1$ & -- & -- \\
\hline
\end{tabular}
\begin{flushleft}
\footnotesize
\justifying
\vspace{-0.2cm}
\hspace{4cm}
$^a$ \cite{Aoki2025}, $^b$ \cite{Roederer2014}
\end{flushleft}
\end{table}

\begin{figure*}[!htb]
    \centering
    \includegraphics[width=1\textwidth]{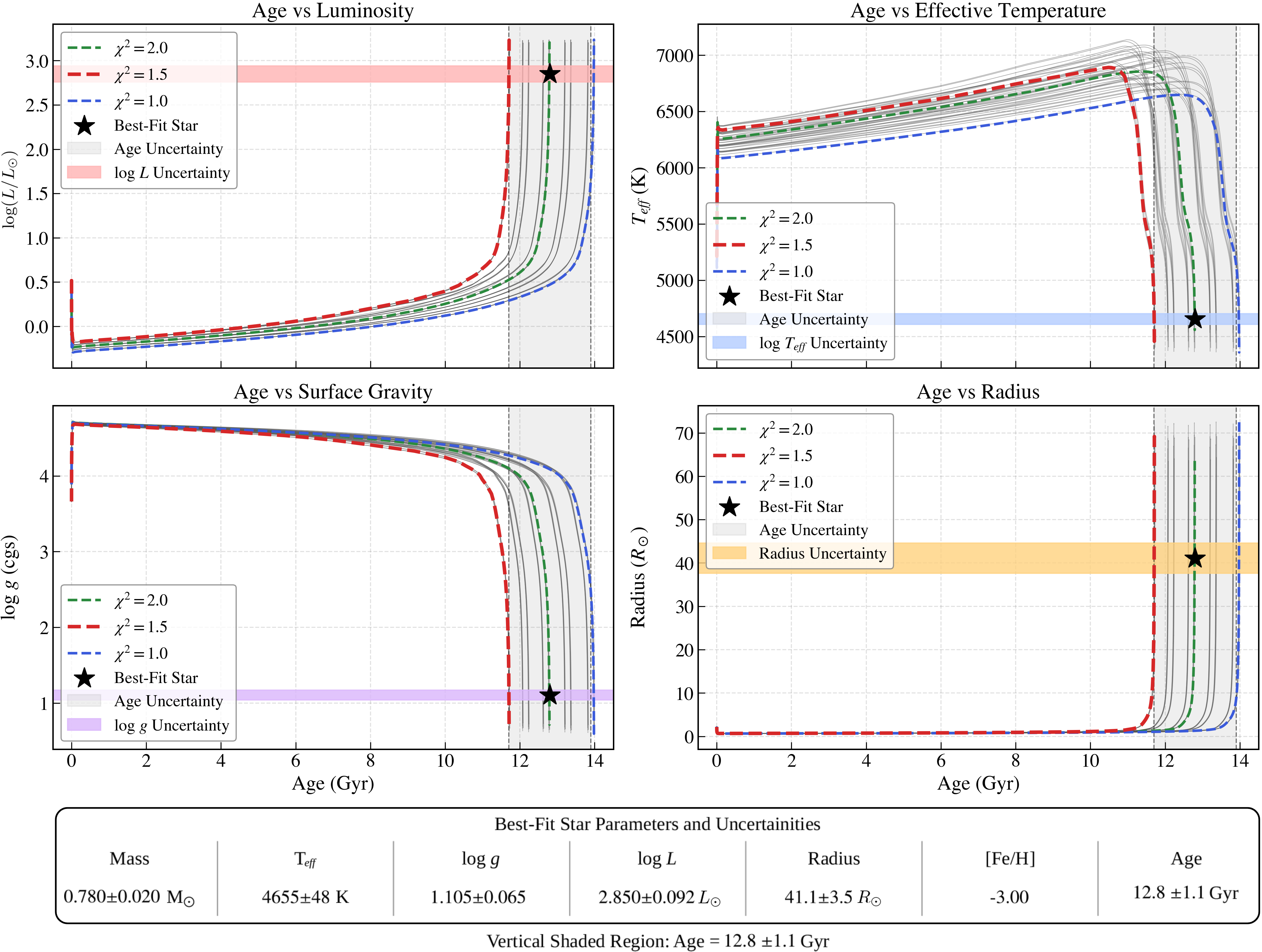}
    \caption{
    Evolutionary constraints on the mass and age of BD$-18^{\circ}$5550. The four panels show the evolution of (top left) luminosity, (top right) effective temperature, (bottom left) surface gravity, and (bottom right) stellar radius as functions of evolutionary age for the calculated stellar-mass grid. Thin grey curves represent individual evolutionary tracks, while the coloured dashed curves indicate representative $\chi^2$ levels. The horizontal shaded bands denote the observational constraints on the corresponding stellar parameters, and the black star marks the adopted best-fitting model. The vertical shaded region represents the inferred age uncertainty. The simultaneous agreement between the evolutionary tracks and the observed luminosity, $T_{eff}$, $\log g$, and radius constrains BD$-18^{\circ}$5550 to an old, low-mass RGB evolutionary state. The figure also illustrates the mass--age degeneracy, whereby models with different initial masses can reproduce similar observed RGB properties at different evolutionary ages.
    }
    \label{fig:evolutionary_tracks}
\end{figure*}

\subsection{Calibration Procedure: Two-Stage Chi-Squared Minimisation}
\label{sec:calibration}
To identify the stellar models that best reproduce the observed properties of BD$-18^\circ$5550, we employ a two-stage $\chi^2$-based calibration procedure. In the first stage, we constrain the fundamental stellar structure using three observables, while in the second stage we refine the stellar physics and nucleosynthetic properties by simultaneously fitting eight observables.

For the fundamental structural calibration, we perform the $\chi^2$ minimization using the three observables that primarily constrain the global stellar properties,
\begin{equation}
\{O_1,O_2,O_3\} = \{T_{eff},\log g,[\mathrm{Fe/H}]\}.
\label{eq:stage1_observables}
\end{equation}
The effective temperature ($T_{eff}$) and surface gravity ($\log g$) constrain the evolutionary state and global stellar structure, while $[\mathrm{Fe/H}]$ constrains the surface metallicity and therefore the chemical composition adopted in the stellar models. We define the reduced $\chi^2$ for this calibration as
\begin{equation}
\chi^2_{\mathrm{red}} = \frac{1}{N_{\rm obs}}
\sum_{i=1}^{3} \left(\frac{O_{i,\mathrm{obs}}-O_{i,\mathrm{model}}} {\sigma_i} \right)^2,
\label{eq:redchi2}
\end{equation}
where $N_{\rm obs}=3$ and $\sigma_i$ represents the observational uncertainty of the $i$th observable. We retain models satisfying $1.0\leq\chi^2_{\mathrm{red}}\leq2.0$ as acceptable structural solutions. This criterion identifies models that provide a simultaneous agreement with the three adopted observational constraints within the explored model grid. From the 1152 models in the structural grid, 45 models satisfy this criterion. These 45 models show a clear clustering in the relevant parameter space, as summarized in Table~\ref{tab:stage1_best_fit}. The concentration of the accepted models in observable space provides the structural constraints used to define the parameter range for the subsequent physics calibration. The evolutionary tracks simultaneously reproduce the observed luminosity, effective temperature, surface gravity, and radius, restricting BD$-18^\circ$5550 to a low-mass, old RGB evolutionary state. Figure~\ref{fig:evolutionary_tracks} shows the evolutionary tracks of the model grid in the observational parameter space of BD$-18^{\circ}$5550. The four panels present the evolution of luminosity, effective temperature, surface gravity, and radius as functions of stellar age. The individual grey curves represent the calculated stellar models, while the coloured dashed curves indicate representative $\chi^2$ levels. The horizontal shaded bands denote the corresponding observational constraints, and the black star marks the adopted best-fitting solution. The tracks remain relatively compact during the main-sequence and early post-main-sequence phases, followed by a rapid change in the stellar properties as the models ascend the RGB. In particular, the luminosity and radius increase rapidly, while the effective temperature and surface gravity decrease as the stellar envelope expands. The observed values of $T_{eff}$, $\log g$, $\log(L/L_\odot)$, and $R/R_\odot$ are simultaneously reproduced only within a restricted region of the evolutionary tracks, thereby constraining the evolutionary state of BD$-18^{\circ}$5550. The figure also illustrates the mass--age degeneracy inherent in the determination of the stellar parameters. Evolutionary tracks with different initial masses can pass through similar regions of the observational parameter space, but at different evolutionary ages. Thus, an individual observable does not provide a unique determination of the stellar mass or age. The simultaneous use of $T_{eff}$, $\log g$, $\log L/L_\odot$, and $R/R_\odot$ significantly reduces this degeneracy. The minimum-$\chi^2$ region therefore provides the preferred combination of initial mass and evolutionary age used in the subsequent stellar-structure and abundance analysis. The complete set of all 45 accepted structural models and their corresponding stellar parameters is provided in Table \ref{tab:calibration_stage1_summary} (Appendix~\ref{appendix:calibration}).

\begin{table*}[!htb]
\centering
\caption{Final stellar and surface-abundance parameters for BD$-18^\circ$5550 from the Stage~2 physics calibration grid. The values presented are the mean and standard deviation from the ensemble of good-fit models from the physics grid (criterion: $1 \le \chi^2_{\rm red} \le 2$), accounting for the full age range uncertainty (11.7--13.9 Gyr). $\Delta$ denotes the difference between the model mean and the adopted observational value, $\Delta = {\rm Model} - {\rm Observed}$. Abundances are expressed in the standard spectroscopic notation [X/Fe] relative to solar photospheric values \citep{Asplund2021}.}
\label{tab:stage2_best_fit}
\begin{tabular}{l l c c c}
\hline
Parameter & Description & Model & Observed & $\Delta$ \\
\hline
$\rm M_{\rm init}/M_{\odot}$ & Initial stellar mass & $0.780\pm0.020$ & --- & --- \\
$\rm M_{\rm current}/M_{\odot}$ & Current stellar mass & $0.756\pm0.016$ & --- & --- \\
$\rm M_{\rm He,core}/M_{\odot}$ & Helium core mass & $0.427\pm0.027$ & --- & --- \\
$T_{eff}$ (K) & Effective temperature & $4660\pm71$ & $4660\pm34$\tablenotemark{a,b} & $+0$ K\\
$\log g$ (cgs) & Surface gravity & $1.149\pm0.157$ & $1.1\pm0.14$\tablenotemark{a,b} & $+0.049$\\
$\log(L/L_{\odot})$ & Bolometric luminosity & $2.796\pm0.198$ & --- & --- \\
$R/R_{\odot}$ & Stellar radius & $38.4\pm10.6$ & --- & --- \\
$[\mathrm{Fe/H}]$ (dex) & Iron metallicity & $-3.00\pm0.05$ & $-3.03\pm0.09$\tablenotemark{a} & $+0.03$ \\
$[\mathrm{C/Fe}]$ (dex) & Carbon-to-iron abundance ratio & $+0.12\pm0.01$ & $-0.02\pm0.04$\tablenotemark{c} & $+0.14$ \\
$[\mathrm{N/Fe}]$ (dex) & Nitrogen-to-iron abundance ratio & $-0.56\pm0.16$ & $-0.40\pm0.15$\tablenotemark{d} & $-0.16$ \\
$[\mathrm{O/Fe}]$ (dex) & Oxygen-to-iron abundance ratio & $+0.57\pm0.12$ & $+0.40\pm0.10$\tablenotemark{e} & $+0.17$ \\
$[\mathrm{Mg/Fe}]$ (dex) & Magnesium-to-iron abundance ratio & $+0.63\pm0.04$ & $+0.58\pm0.11$\tablenotemark{a} & $+0.05$ \\
$[\mathrm{C/N}]$ (dex) & Carbon-to-nitrogen ratio (relative to solar) & $+0.69\pm0.16$ & $+0.34$\tablenotemark{d} & $+0.35$ \\
$^{12}\mathrm{C}/^{13}\mathrm{C}$ & Carbon isotopic ratio (number) & $41.9\pm5.7$ & $>40$\tablenotemark{d} & --- \\
$^{14}\mathrm{N}/^{15}\mathrm{N}$ & Nitrogen isotopic ratio (number) & $60.3\pm22.1$ & --- & --- \\
$^{16}\mathrm{O}/^{18}\mathrm{O}$ & Oxygen isotopic ratio (number) & $1400\pm370$ & --- & --- \\
Age (Gyr) & Evolutionary age & $11.87\pm0.16$\tablenotemark{f} & --- & --- \\
$\alpha_{\rm MLT}$ & Mixing-length parameter & $1.9\pm0.2$ & --- & --- \\
$\alpha_{\rm th}$ & Thermohaline efficiency & $1.0\pm0.5$ & --- & --- \\
$f_{\rm N}$ & Nitrogen reaction rate enhancement factor & $0.004\pm0.001$ & --- & --- \\
$f_{\rm Mg}$ & Magnesium opacity enhancement factor & $0.8\pm0.2$ & --- & --- \\
$[\alpha/\mathrm{Fe}]$ & Overall $\alpha$-element enhancement & $0.4\pm0.1$ & --- & --- \\
\hline
\end{tabular}
\begin{flushleft}
\footnotesize
\justifying
\vspace{-0.2cm}
$^a$ \cite{Aoki2025}, $^b$ \cite{Roederer2014}, $^c$ \cite{Masseron2010}, $^d$ \cite{Spite2005,Spite2006}, $^e$ \cite{Cayrel2004}, $^f$ Stage 1 structural calibration yielded Age $= 12.8 \pm 1.1$ Gyr. However, Stage 2 physics calibration, incorporating elemental abundances and isotope ratios, reveals a tighter constraint of $11.87\pm0.16$ Gyr. This younger age reflects the powerful constraints of CNO nucleosynthesis abundances and $^{12}$C/$^{13}$C ratios on deep mixing, and remains within the Stage 1 uncertainty. The age shift demonstrates that abundance ratios provide superior age precision for metal-poor RGB stars.
\end{flushleft}
\end{table*}

Having constrained the stellar structure in Stage~1, we now construct a finer Physics Model Grid designed to simultaneously reproduce eight observables sensitive to the adopted stellar physics, chemical composition, nucleosynthesis, and mixing processes. We perform the
$\chi^2$ minimization using
\begin{equation}
\{O_1,O_2,\ldots,O_8\} = \left\{T_{eff},\log g,[\mathrm{Fe/H}], [\mathrm{C/Fe}],[\mathrm{N/Fe}], [\mathrm{O/Fe}],[\mathrm{Mg/Fe}], ^{12}\mathrm{C}/^{13}\mathrm{C} \right\}.
\label{eq:stage2_observables}
\end{equation}
We retain the first three observables, $T_{eff}$, $\log g$, and $[\mathrm{Fe/H}]$, from Stage~1 to ensure that the models satisfying the additional abundance constraints remain consistent with the previously established structural solution. The reduced $\chi^2$ for Stage~2 is calculated using Equation~\ref{eq:redchi2}, with $N_{\rm obs}=8$. We again select models satisfying $1\leq\chi^2_{\mathrm{red}}\leq2$ as acceptable solutions. From the 1152 models in the Physics Grid, exactly 12 models satisfy this selection criterion. These 12 good-fit models characterized by the adopted values of $\alpha_{\rm MLT}$, $\alpha_{\rm TH}$, $f_{\rm N}$, $f_{\rm Mg}$, and $[\alpha/\mathrm{Fe}]$, as summarized in Table~\ref{tab:stage2_best_fit}. The complete set of all 12 accepted physics models and their corresponding stellar parameters is provided in Table \ref{tab:calibration_stage2_summary} (Appendix~\ref{appendix:calibration}).
The inclusion of five additional observables sensitive to nucleosynthesis and mixing dramatically constrains the stellar age and mass. Whereas Stage~1 (using structural constraints alone) yielded $\rm M = (0.780 \pm 0.020)\,M_{\odot}$ and $t = (12.8 \pm 1.1)\,\text{Gyr}$, the Stage~2 analysis refines these to $\rm M = (0.756 \pm 0.016)\,M_{\odot}$ and $t = (11.87 \pm 0.16)\,\text{Gyr}$. The age uncertainty is improved by a factor of $\sim 7$ (from $\pm 1.1$ Gyr to $\pm 0.16$ Gyr), reflecting the strong constraint provided by CNO abundances and isotope ratios on the CNO cycle progress and thus the core temperature history. The small shift in age ($\Delta t \approx -0.9$ Gyr) is driven primarily by the constraint from $^{12}$C/$^{13}$C, which indicates that the CNO cycle has progressed further than predicted by the Stage~1 age, requiring a slightly younger (more evolved) model.

\begin{figure*}[!htb]
    \centering
    \includegraphics[width=1\textwidth]{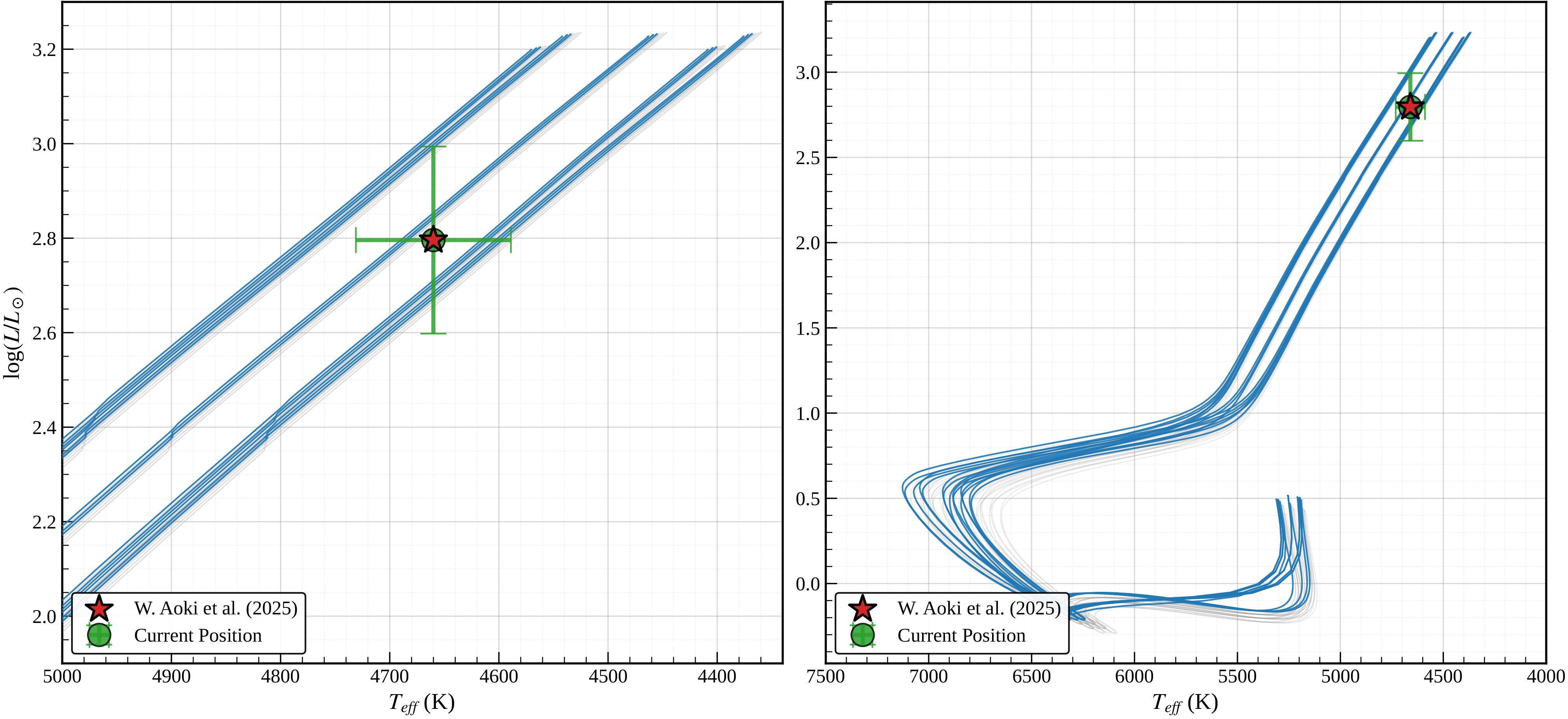}
    \caption{Hertzsprung--Russell diagram showing the evolutionary tracks calculated for BD$-18^{\circ}$5550. The right panel presents the full evolutionary tracks over the calculated evolutionary sequence, while the left panel shows an enlarged view of the region around the observed stellar position. The green circle represents the observed effective temperature and luminosity, with the associated uncertainties shown by the error bars. The red star marks the position of the adopted model at the selected evolutionary state. The agreement between the model and observed position in the enlarged panel demonstrates that the adopted evolutionary model reproduces the observed $T_{eff}$ and luminosity simultaneously and places BD$-18^{\circ}$5550 on the ascending RGB.
    }
    \label{fig:hr_diagram}
\end{figure*}

\subsection{Evolutionary Location in the Hertzsprung--Russell Diagram}
\label{sec:hrdiagram}
The evolutionary location of BD$-18^{\circ}$5550 was further examined in the Hertzsprung--Russell (HR) diagram by comparing the calculated evolutionary tracks with the observed effective temperature and luminosity. Figure~\ref{fig:hr_diagram} shows the resulting $T_{eff}$--$\log(L/L_\odot)$ diagram. The right-hand panel presents the full evolutionary tracks, while the left-hand panel provides an enlarged view of the region occupied by the observed star. The green circle denotes the current position of BD$-18^{\circ}$5550, with the horizontal and vertical error bars representing the corresponding uncertainties in $T_{eff}$ and luminosity, respectively. The red star indicates the position of the star adopted from literature \citep{Aoki2025}. The close correspondence between the model position and the observed constraints demonstrates that the adopted model simultaneously reproduces the observed effective temperature and luminosity.

The full evolutionary tracks shown in the right-hand panel illustrate the evolution from the ZAMS phases to the RGB. During the main-sequence and subsequent post-main-sequence evolution, the models occupy a relatively confined region of the HR diagram before moving toward lower effective temperatures and rapidly increasing luminosities during RGB ascent. The observed position of BD$-18^{\circ}$5550 is located on the ascending RGB, indicating that the star is in an advanced giant evolutionary phase. The different tracks also demonstrate the dependence of the evolutionary path on the initial stellar mass, with models of different masses reaching similar regions of the HR diagram at different evolutionary times. The enlarged view in the left panel provides a more direct comparison between the model and the observations. The adopted model lies within the observational uncertainties in the $T_{eff}$--luminosity plane, providing an independent validation of the evolutionary solution obtained from the simultaneous constraints on $T_{eff}$, $\log g$, luminosity, and radius. 

The convergence of evolutionary tracks near the RGB tip is evident in both panels,  reflecting the rapid evolution that low-mass stars experience on the upper RGB as the convective envelope deepens and the core contracts. The timescale for core helium burning at this evolutionary stage is only $\sim 1-2\,\text{Gyr}$, explaining why high-precision age determinations are possible for RGB stars, small changes in age produce measurable changes in stellar parameters.

\begin{figure*}[!htb]
    \centering
    \includegraphics[width=1\textwidth]{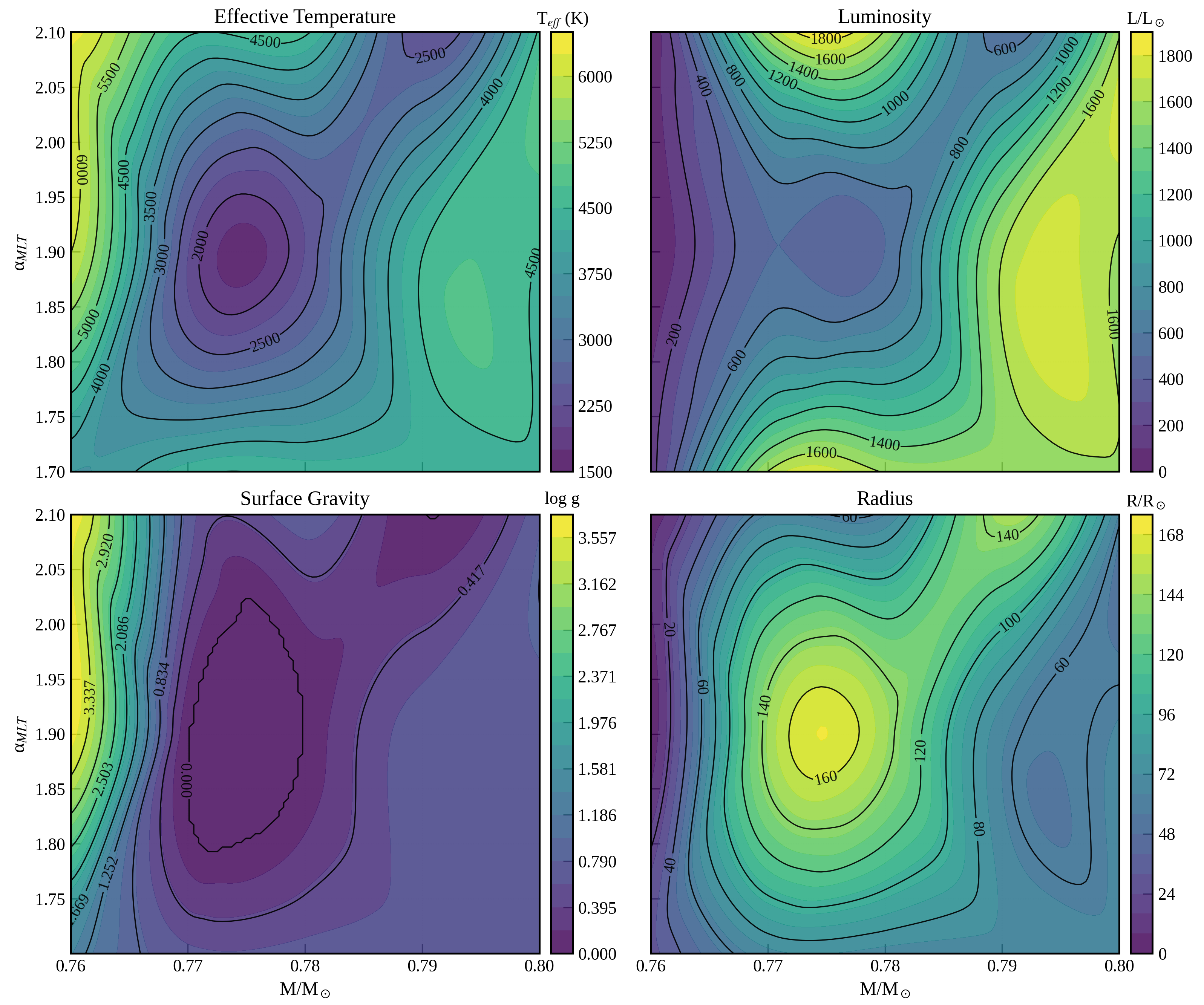}
    \caption{Sensitivity of the stellar properties of BD$-18^{\circ}$5550 to the initial stellar mass and mixing-length parameter. The four panels show the distribution of (a) effective temperature, (b) luminosity, (c) surface gravity, and (d) stellar radius in the $(M_{\rm init},\alpha_{\rm MLT})$ parameter space. The colour scale and contours indicate the corresponding model predictions. The results demonstrate that the stellar observables depend simultaneously on the initial mass and the efficiency of convective energy transport. In particular, similar values of $\log g$ can be obtained for different combinations of $M_{\rm init}$ and $\alpha_{\rm MLT}$, illustrating the mass--convection degeneracy. The combined constraints from $T_{eff}$, luminosity, $\log g$, and radius are therefore required to identify the preferred region of parameter space.}
    \label{fig:mass_alpha}
\end{figure*}

\subsection{Sensitivity to Stellar Mass and Convective Efficiency}
\label{sec:mass_alpha_sensitivity}
To investigate the sensitivity of the inferred stellar properties to the initial stellar mass and the efficiency of convective energy transport, we performed a two-dimensional parameter study in the $\rm (M_{init},\alpha_{\rm MLT})$ plane. Figure~\ref{fig:mass_alpha} presents the distributions of the effective temperature, luminosity, surface gravity, and stellar radius across the explored parameter space. The resulting contours demonstrate that the stellar properties depend on $\rm M_{init}$ and $\alpha_{\rm MLT}$ in a strongly coupled and non-linear manner. The $T_{eff}$ distribution (top left) exhibits substantial variation across the parameter space, with a pronounced low-temperature region near $\rm M_{\rm init}\simeq0.775\,M_{\odot}$ and $\alpha_{\rm MLT}\simeq1.9$. The luminosity distribution (top right) increases systematically with both $\rm M_{\rm init}$ and $\alpha_{\rm MLT}$, with the brightest luminosities achieved at high values of both parameters. The surface gravity (bottom left) varies strongly across the grid and is particularly sensitive to the changes in stellar radius. This behaviour follows from the relation $g=G\rm{M}/R^2$, such that variations in radius can dominate over the relatively small changes in stellar mass. The radius distribution (bottom right) shows a broad maximum around $\rm M_{\rm init}\simeq0.775\,M_{\odot}$ and $\alpha_{\rm MLT}\simeq1.9$, with substantially smaller radii toward other regions of the parameter space.

The combined behaviour of these four quantities demonstrates that neither $\rm M_{\rm init}$ nor $\alpha_{\rm MLT}$ can be constrained reliably from a single observable. We therefore use the available observational constraints from the literature, namely $T_{eff}$, $\log g$, and $[\mathrm{Fe/H}]$, to identify the region of the $\rm(M_{\rm init},\alpha_{\rm MLT})$ parameter space that is consistent with the observed stellar properties. The evolutionary tracks in the Figures~\ref{fig:evolutionary_tracks} and \ref{fig:hr_diagram} provide complementary tests of the evolutionary models and illustrate how the observed effective temperature and surface gravity constrain the evolutionary sequences. The adopted metallicity, constrained by the observed $[\mathrm{Fe/H}]$, defines the chemical composition of the model grid used in the evolutionary calculations. This combined sensitivity analysis provides the basis for selecting the adopted mixing-length parameter and refining the initial-mass range used in the subsequent evolutionary calculations.

\begin{figure*}[!htb]
    \centering
    \includegraphics[width=1\textwidth]{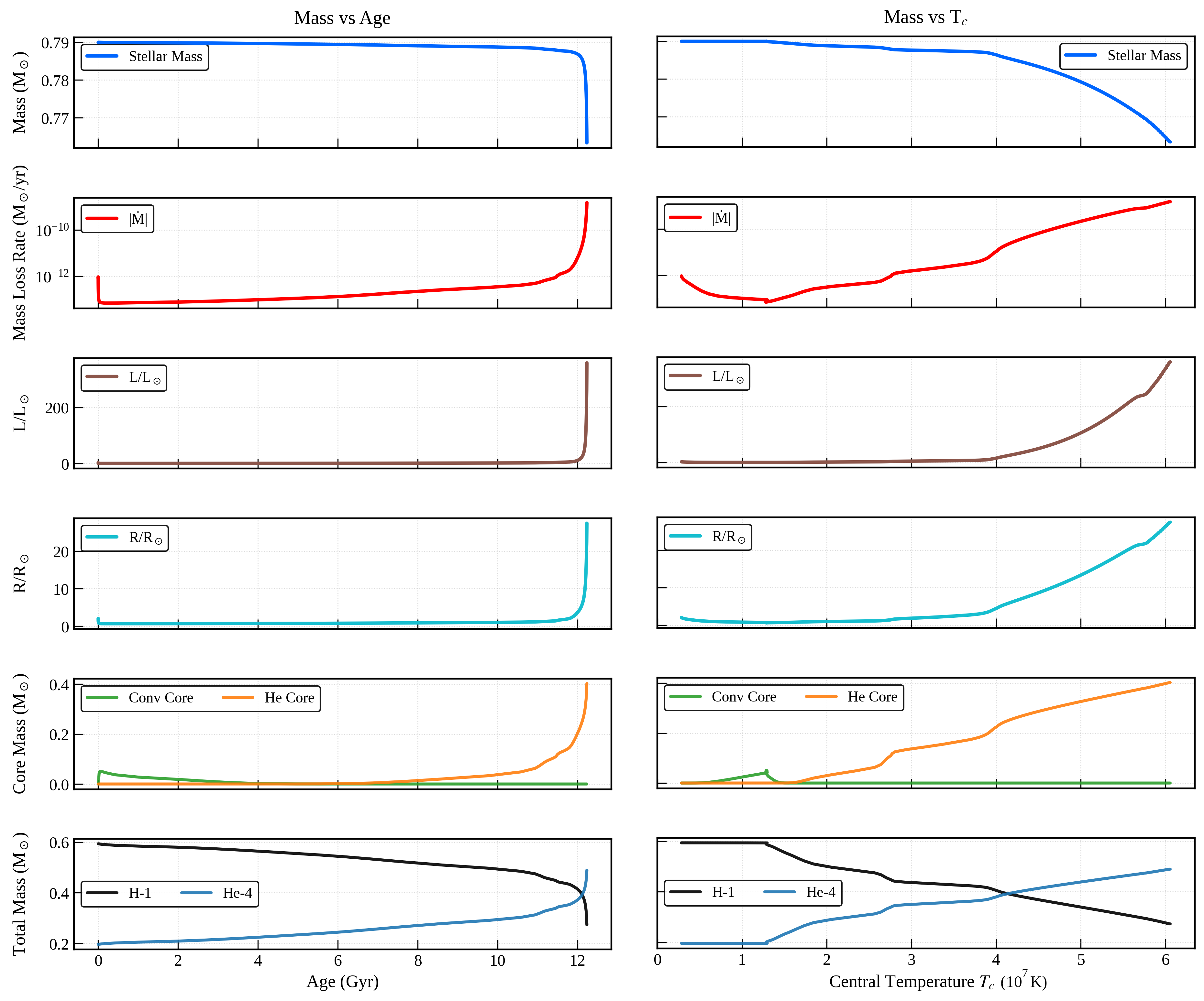}   
    \caption{Structural and mass-loss evolution of the adopted model of BD$-18^{\circ}$5550. The left column shows the evolution as a function of stellar age, while the right column shows the same quantities as a function of central temperature, $T_{\rm c}$. From top to bottom, the panels show the total stellar mass, absolute mass-loss rate, luminosity, stellar radius, convective and helium-core masses, and hydrogen and helium mass fractions.}
    \label{fig:mass_loss_evolution}
\end{figure*}

\section{Mass Loss, Stellar Structure, and Core Evolution}
\label{sec:mass_loss_structure}
The evolution of stellar mass, luminosity, radius, and internal structure of the evolutionary phase is illustrated in Figure~\ref{fig:mass_loss_evolution}. These panels collectively demonstrate how low-mass stars transform from main-sequence to giant with radii exceeding $40\,R_{\odot}$ over a timescale comparable to the age of the Galaxy.

The stellar mass decreases from an initial value of $\rm M_{\rm init} = (0.780 \pm 0.020)\,M_{\odot}$ to a present-day value of $\rm M_{\rm current} = (0.756 \pm 0.016)\,M_{\odot}$ due to wind-driven mass loss, resulting in a net loss of $\rm \Delta M \approx 0.024\,M_{\odot}$ over $11.87\,\mathrm{Gyr}$. This mass loss follows the Reimers prescription with a mass-loss coefficient $\eta = 0.3$, calibrated to match the observed white-dwarf mass distribution in galactic clusters \citep{Reimers1975, Kalirai2008}. The mass loss is negligible during the main sequence and early RGB, but increases dramatically in the final $1-2\,\mathrm{Gyr}$ as the star expands and its luminosity increases. The model retains approximately $97\%$ of its initial mass for most of its lifetime, with the majority of mass loss occurring only in the advanced RGB phase when the mass-loss rate reaches values of order $\rm \sim 10^{-10}\,M_{\odot}\,\mathrm{yr}^{-1}$.

The evolution of stellar luminosity demonstrates the dramatic transformation of the star from main sequence to the current RGB phase. On the ZAMS, the luminosity was $\log(L/L_{\odot}) = 0.520$, corresponding to $L = 3.3\,L_{\odot}$—a moderately luminous star even at birth. As the star evolved through the main sequence via core hydrogen burning, the luminosity increased gradually to $\log(L/L_{\odot}) = 0.791$ (approximately $L = 6.2\,L_{\odot}$) at core hydrogen exhaustion around age $\sim 10\,\mathrm{Gyr}$, representing a modest 1.9-fold increase over the $\sim 10\,\mathrm{Gyr}$ main-sequence lifetime. Following the onset of hydrogen-shell burning on the RGB, the luminosity increased dramatically, reaching the current value of $\log(L/L_{\odot}) = (2.796 \pm 0.198)$ at age $11.87\,\mathrm{Gyr}$, corresponding to $L = (625 \pm 295)\,L_{\odot}$. This represents a spectacular $\sim 101$-fold increase from core hydrogen exhaustion to the present day, accomplished in just $\sim 1.87\,\mathrm{Gyr}$. Over the entire stellar lifetime from ZAMS to present, the luminosity has increased by a factor of $\sim 189$. This dramatic rise during the RGB phase reflects the increasing power output of the H-burning shell as the core contracts and heats, providing the energy budget needed to support the giant radius and sustain rapid expansion.

The stellar radius undergoes dramatic expansion from $\sim 0.5\,R_{\odot}$ on the main sequence to $R = 38.4 \pm 10.7\,R_{\odot}$ on the current upper RGB—a 70-fold increase. Most of this expansion occurs during the RGB phase (age $> 10\,\mathrm{Gyr}$), when the deepening convective envelope causes the star to cool and expand to maintain hydrostatic pressure balance. The expansion is particularly rapid during the final $1-2\,\mathrm{Gyr}$, with the radius increasing by a factor of $\sim 20$ during this short interval. This rapid expansion is one of the hallmarks of RGB evolution and makes RGB stars easily identifiable in colour-magnitude diagrams.

The helium-core mass increases steadily from $\rm \sim 0.12\,M_{\odot}$ at age $11\,\mathrm{Gyr}$ (near the onset of the upper RGB) to $\rm M_{\rm He,core} = (0.427 \pm 0.027)\,M_{\odot}$ at the current age of $11.87\,\mathrm{Gyr}$. This rapid growth of the helium core in the final $\sim 0.9\,\mathrm{Gyr}$ reflects the greatly increased energy output from the hydrogen-burning shell as the core contracts and heats, causing hydrogen burning to proceed rapidly and produce helium ash at an accelerating rate. At the core, hydrogen has been completely exhausted ($X = 0.00$) and converted to helium ($Y = 1.00$), establishing a dense helium core approaching the conditions for helium ignition. The width of the hydrogen-burning shell remains thin (roughly $\rm 0.02\,M_{\odot}$), concentrating the nuclear energy release in a narrow region near the helium core.
 
The composition of the stellar envelope reveals the evolutionary history of the star. At the ZAMS, the envelope had initial abundances of $X = 0.750$ and $Y = 0.249$. The current-day surface composition remains nearly unchanged at $X = 0.756$ and $Y = 0.243$, reflecting the fact that the main-sequence lifetime vastly exceeds the RGB lifetime. In contrast, the light-element abundances show differential behaviour. Carbon and oxygen remain essentially pristine, while nitrogen has gradually enriched from $[\text{N/Fe}] \approx -2.0\,\text{dex}$ at the ZAMS to $[\text{N/Fe}] \approx -0.56\,\text{dex}$ at the current age, a signature of early-stage thermohaline mixing at the base of the convective envelope. This combined pattern—pristine H and He coupled with pristine C and O but enriching N—confirms that the convective envelope has not penetrated to depths where significant hydrogen or carbon-oxygen nuclear processing has occurred, while thermohaline mixing has modestly altered the nitrogen abundance. The envelope composition thus provides tight constraints on the depth of the convective mixing zone, the efficiency of thermohaline mixing operating during the RGB phase, and the role of mass loss in shaping the star's current structure and abundance profile.
  
The right column of Figure~\ref{fig:mass_loss_evolution} displays the same quantities as functions of central temperature, emphasising the connection between core evolution and observable properties. The central temperature evolves from $T_c \approx 5 \times 10^6\,\mathrm{K}$ on the ZAMS to $T_c \approx 10^8\,\mathrm{K}$ at the current epoch. The sharp increase in luminosity (from $\log L \sim 0.8$ to $\log L > 2.5$) above $T_c \sim 10^7\,\mathrm{K}$ marks the transition from the main sequence to the RGB, where the core has become fully convective and exhausted its hydrogen fuel. The sharp increase in radius above $T_c \sim 5 \times 10^7\,\mathrm{K}$ marks the onset of rapid RGB expansion as the convective envelope grows in response to the contracting, heating core. The steady increase in helium-core mass above $T_c \sim 10^8\,\mathrm{K}$ illustrates the continued accumulation of helium ash in the core from hydrogen-shell burning as the core density and temperature increase toward the conditions required for helium ignition.

\section{Abundance Estimation}
\label{sec:abundances}
The surface elemental abundance pattern of BD\,-\,18\textdegree5550 reveals a classic metal-poor RGB star with enhanced $\alpha$-element abundances characteristic of EMP populations in the Galactic halo.  The abundances were computed using the standard logarithmic scale defined by the equation $[\mathrm{X/Fe}] = \log ({N_X}/{N_{\mathrm{Fe}}})_{\rm *} - \log ( {N_X}/{N_{\mathrm{Fe}}})_{\odot}$. The solar reference values are taken from \cite{Asplund2021}. Table~\ref{tab:elemental_abundances} presents the modelled surface abundances compared to observational constraints. The complete isotopic surface abundance output from the extended nuclear network is presented in Table~\ref{tab:isotopes_fe} (Appendix \ref{appendix:abundances}). 

\begin{table}[ht]
\centering
\caption{Surface mass fractions and elemental abundances of stable elements in the best-fit BD$-18^\circ$5550 model at an age of  11.87~Gyr, compared with observed values, $\rm \Delta=Model - Observed$. Only astrophysically relevant and observationally accessible species are listed. Abundances are expressed in the standard spectroscopic notation [X/Fe] relative to solar photospheric values \citep{Asplund2021}. The complete surface abundance output from the extended nuclear network is presented in Table~\ref{tab:isotopes_fe}.}
\label{tab:elemental_abundances}
\begin{tabular}{lcccc}
\hline
Element &
$[\mathrm{X/Fe}]_{\rm model}$ & $[\mathrm{X/Fe}]_{\rm observed}$ &
$\Delta$ & Reference \\
\hline
H  & --- & --- & --- & --- \\
He & --- & --- & --- & ---\\
C  &  $+0.12\pm0.01$ & $-0.02\pm0.15$ & $+0.14$ & \cite{Masseron2010} \\
N & $-0.56\pm0.16$ & $-0.40\pm0.15$ & $-0.16$ & \cite{Spite2005,Spite2006} \\
O  & $+0.57\pm0.12$ & $+0.40\pm0.10$ & $+0.17$ & \cite{Cayrel2004}\\
Ne &  $+0.54\pm0.12$ & --- & --- & --- \\
Na &  $+0.16\pm0.05$ & --- & --- & --- \\
Mg &  $+0.63\pm0.04$ & $+0.58\pm0.11$ & $+0.05$ & \cite{Aoki2025} \\
Al &  $+0.14\pm0.01$ & --- & --- & --- \\
Si &  $+0.41\pm0.15$ & --- & --- & ---\\
S  &  $+0.42\pm0.10$ & --- & --- & ---\\
Ar &  $+0.07\pm0.01$ & --- & --- & ---\\
Ca &  $+0.43\pm0.08$ & --- & --- & ---\\
\hline
\end{tabular}
\end{table}
 
\begin{figure*}[!htb] 
    \centering 
    \includegraphics[width=1\textwidth]{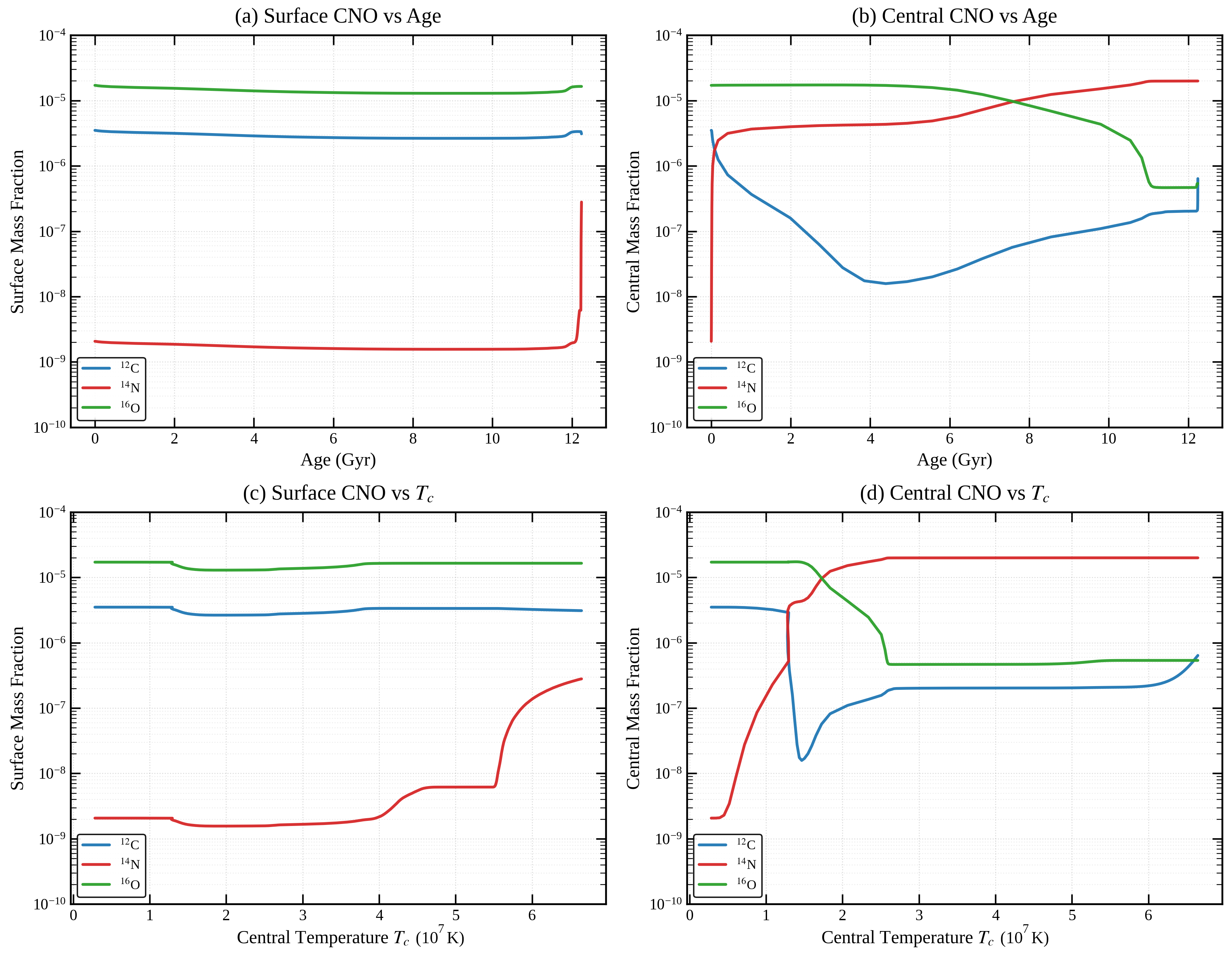} 
    \caption{Evolution of CNO abundances in the $\rm M_{init}=0.78\,M_{\odot}$ model of BD$-18^\circ$5550. Panels (a) and (b) show the surface and central mass fractions of $^{12}$C, $^{14}$N, and $^{16}$O as functions of evolutionary age, while panels (c) and (d) show the corresponding abundances as functions of central temperature $T_{c}$.} 
    \label{fig:cno_evolution} 
\end{figure*}

\subsection{CNO Nucleosynthesis and First Dredge-Up}
\label{sec:cno_evolution}
The evolution of carbon, nitrogen, and oxygen abundances during the RGB phase provides crucial diagnostics of mixing processes and nuclear burning rates. Figure~\ref{fig:cno_evolution} displays the mass-fraction evolution of these elements from the pre-main sequence through the current age of $11.87\,\text{Gyr}$. The left panels of Figure~\ref{fig:cno_evolution} show surface abundances, which are directly comparable to spectroscopic observations. A crucial finding from our models is that the surface carbon and oxygen abundances remain essentially pristine throughout the entire evolution up to the current age, while the nitrogen abundance shows evidence of early-stage enrichment consistent with the onset of thermohaline mixing. The surface $^{12}$C and $^{16}$O abundances show negligible change from the ZAMS ($t = 0$) through the current epoch ($t = 11.87\,\text{Gyr}$). This remarkable constancy reflects the fact that the convective envelope, even as it deepens during the RGB phase, does not yet penetrate to the depths where significant nuclear processing of carbon and oxygen has occurred via the CNO cycle. In contrast, the surface $^{14}$N abundance exhibits a gradual increase over time (Panel~a), rising from approximately $[\text{N/Fe}] \approx -2.0$ dex at the ZAMS to $[\text{N/Fe}] \approx -0.4$ dex at the current age. This increase in nitrogen, though modest and coupled with unchanged carbon and oxygen, indicates that early-stage thermohaline (salt-fingering) mixing has begun to operate at the base of the convective envelope, slowly introducing nitrogen-enriched material processed by the CNO cycle in earlier phases into the surface layers. The surface $\rm [C/N]$ number ratio remains approximately $\rm +0.69\,dex$, reflecting the constancy of carbon combined with the modest increase in nitrogen. This intermediate state—where carbon and oxygen remain pristine, but nitrogen shows gentle enrichment—provides a powerful diagnostic that the star is at an intermediate evolutionary stage on the RGB, after the ZAMS but prior to the onset of the dramatic first dredge-up event that would alter all three light-element abundances significantly. The persistence of high [C/Fe] combined with the rising [N/Fe] confirms that first dredge-up has not yet occurred, which would place the star at a more advanced evolutionary stage where carbon abundance would be substantially reduced and nitrogen dramatically enhanced.

The right panels of Figure~\ref{fig:cno_evolution} display central abundances, which reveal the ongoing nuclear burning in the core and shell. The central $^{12}\text{C}$ abundance decreases steeply with age and central temperature due to the $\alpha$-capture reaction $^{12}\text{C}(\alpha,\gamma)^{16}\text{O}$, which operates at core temperatures exceeding $T_c \sim 10^7\,\text{K}$. At the current epoch, the central $^{12}\text{C}$ abundance is depleted to extremely low levels ($< 10^{-8}$ mass fraction), indicating that nearly all carbon initially present in the core has been converted via this reaction. The central $^{14}\text{N}$ abundance, by contrast, increases gradually at intermediate temperatures and then begins to decrease at the highest central temperatures ($T_c > 10^8\,\text{K}$), reflecting the transition from production via the CNO cycle (which converts $^{12}\text{C}$ to $^{14}\text{N}$) to destruction via $^{14}\text{N}(\alpha,\gamma)^{18}\text{O}$ (which consumes $^{14}\text{N}$ to produce oxygen). The central $^{16}\text{O}$ abundance increases dramatically with temperature due to the $^{12}\text{C}(\alpha,\gamma)^{16}\text{O}$ reaction, providing a reservoir of oxygen for alpha-capture reactions. At the current central temperature ($T_c \sim 10^8\,\text{K}$), the central oxygen abundance is elevated to approximately $10^{-5}$ mass fraction. The ongoing accumulation of central oxygen is significant because it presages the triple-alpha process and helium ignition, which will occur once the core temperature reaches $T_c \sim 10^8\,\text{K}$ and the core mass accumulates sufficient helium.

\begin{figure*}[!htb]
    \centering
    \includegraphics[width=1\textwidth]{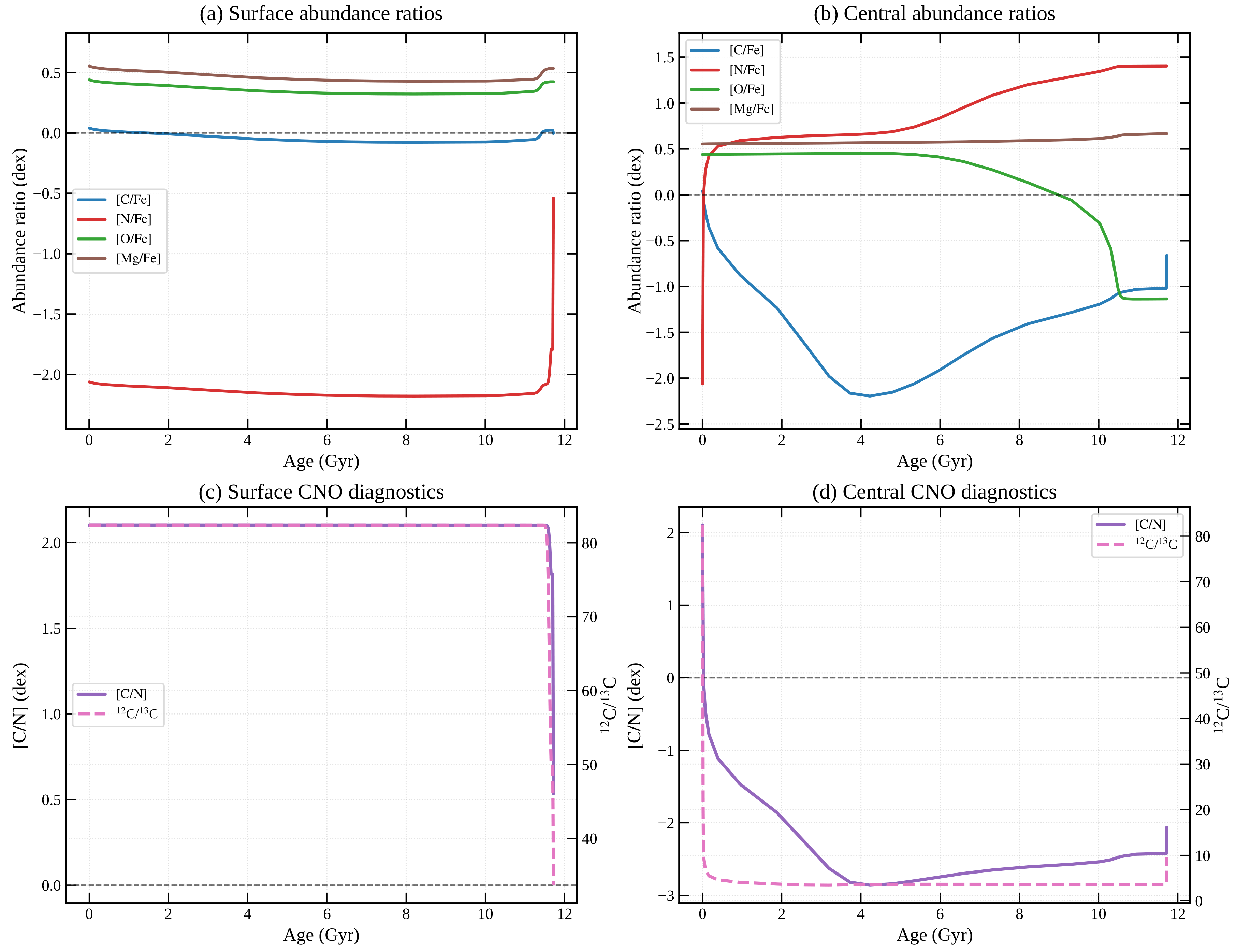}   
    \caption{Abundance-Ratio Diagnostics of BD$-18^\circ$5550. The left column shows the surface elemental abundance ratios [X/Fe] and CNO diagnostics, whereas the right column shows the central elemental abundance ratios [X/Fe] and CNO diagnostics as functions of stellar age.}
    \label{fig:abundance_ratio}
\end{figure*}

An important diagnostic is the central isotopic ratio $^{12}\text{C}/^{13}\text{C}$, which provides a sensitive measure of the degree to which the CNO cycle has reached equilibrium. Panel (d) of Figure~\ref{fig:abundance_ratio} shows that the central $^{12}\text{C}/^{13}\text{C}$ ratio remains very large ($\gtrsim 6$) at the current age, indicating that the CNO cycle has not yet driven this ratio to its equilibrium value of $\sim 3.7$ (which would require higher central temperatures or much longer equilibration timescales). In contrast, the surface $^{12}\text{C}/^{13}\text{C}$ ratio is predicted by our models to remain $\gtrsim 40$ due to the pristine envelope composition, in excellent agreement with the observational constraint from detailed spectroscopy. This large surface ratio reflects the primordial composition and the absence of significant CNO-cycle processing at the stellar surface. Such a measurement, if obtained through high-resolution spectroscopy, would provide an independent and powerful test of our stellar evolution models and mixing 
prescriptions.

\begin{figure*}[!htb]
    \centering
    \includegraphics[width=1\textwidth]{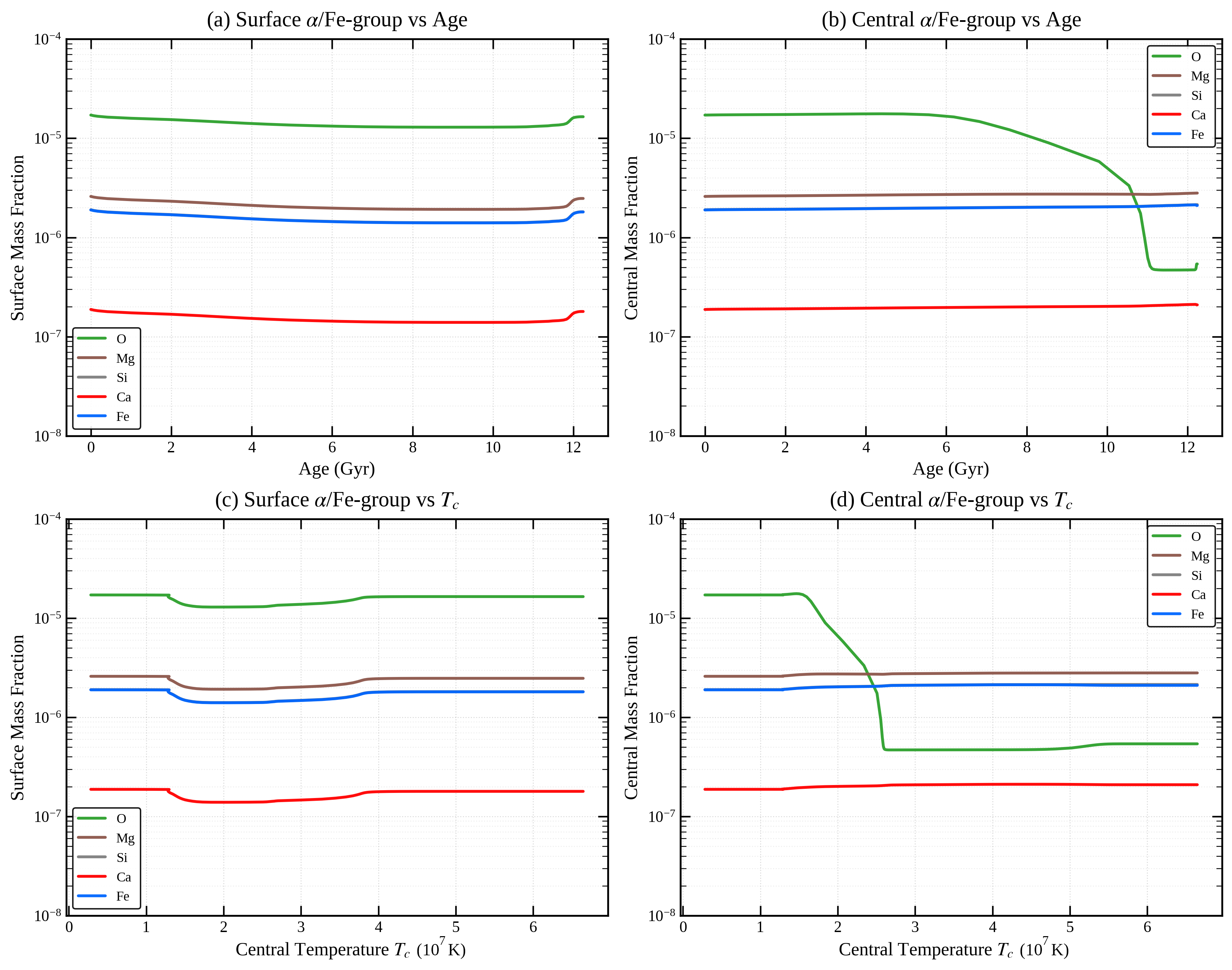}   
    \caption{Evolution of $\alpha$-capture element abundances during the RGB phase of BD$-18^\circ$5550. Left panels (a, c) display surface mass fractions; right panels (b, d) show central mass fractions. Top panels (a, b) show evolution as functions of stellar age; bottom panels (c, d) show the same 
    quantities as functions of central temperature $T_c$.}
    \label{fig:alpha_evolution}
\end{figure*}

\subsection{Alpha-Element Abundances and the Primordial Composition}
\label{sec:alpha_evolution}
The $\alpha$-capture elements (oxygen, magnesium, silicon, and calcium) are produced primarily in core-collapse supernovae of massive stars and are the dominant heavy elements in metal-poor galactic environments. These elements provide diagnostics of the primordial nucleosynthesis in the early Galaxy and constraints on the mixing properties of evolved stars. Figure~\ref{fig:alpha_evolution} displays the evolution of mass fractions for oxygen, magnesium, silicon, calcium, and iron during the stellar evolution of BD$-18^\circ$5550 from the pre-main sequence through the current age of $11.87\,\text{Gyr}$. Panels~(a) and (c) display the surface mass fractions of $\alpha$-capture elements. A striking feature is the remarkable constancy of these abundances across the entire evolutionary sequence. The surface oxygen mass fraction (green line, panel a) remains essentially unchanged at approximately $3.260 \times 10^{-6}$ from the ZAMS ($t = 0$) through the current age ($t = 11.87\,\text{Gyr}$). Similarly, the surface magnesium mass fraction (blue line) stays constant at $\sim 2.958 \times 10^{-6}$, and the surface calcium mass fraction (orange line) remains at $\sim 1.765 \times 10^{-7}$. Variations in all elements are less than $0.05$ dex over the entire evolution. This constancy is even more evident in panel~(c), where surface abundances are plotted against central temperature. As the central temperature increases from $T_c < 10^7\,\text{K}$ (on the main sequence) to $T_c \approx 10^8\,\text{K}$ (on the upper RGB), the surface abundances remain essentially independent of $T_c$. This independence demonstrates that the surface composition is completely decoupled from the nuclear burning occurring in the stellar interior and has not been mixed with core material that has undergone significant nucleosynthetic processing. This behaviour indicates that the convective envelope has not yet deepened to the point where it reaches hydrogen-burning material, confirming that the surface abundances are locked in at their initial values, set by the primordial composition of the interstellar medium at the time of star formation approximately, $z \sim 15-20$ \citep{Searle1978, Iben2012}. The surface abundances thus provide a fossil record of the early Galactic nucleosynthesis and the metal content of the star-forming environment in the 
early Galaxy.

Panels~(b) and (d) of Figure~\ref{fig:alpha_evolution} reveal a dramatically different story for the central abundances. Unlike the surface, the central composition undergoes substantial changes as the star evolves and the core temperature rises. Panel~(b) shows central abundances as functions of age. At early ages ($ t < 2\,\text{Gyr}$), when central temperatures remain below $T_c \sim 10^7\,\text{K}$, all central abundances remain frozen at their initial values, identical to the surface. However, as the core temperature increases with advancing age, the central composition begins to evolve. Most dramatically, the central oxygen mass fraction decreases from an initial value of $\sim 2.15 \times 10^{-5}$ to approximately $5.68 \times 10^{-7}$ by age $11.87\,\text{Gyr}$. This dramatic depletion reflects consumption of oxygen via the triple alpha process ($3\alpha \to ^{12}$C) at intermediate temperatures ($T_c \sim 10^7 - 10^8\,\text{K}$) and via the $^{12}$C$(\alpha,\gamma)^{16}$O reaction at higher temperatures. Panel~(d) makes clear the temperature dependence of these processes. At $T_c < 3 \times 10^7\,\text{K}$, all central abundances remain constant. As $T_c$ increases to $\sim 5 \times 10^7\,\text{K}$, the central oxygen begins to decrease. At $T_c > 7 \times 10^7\,\text{K}$ (corresponding to the upper RGB, age $> 11\,\text{Gyr}$), the oxygen depletion accelerates and the central oxygen mass fraction drops sharply. By $T_c \approx 10^8\,\text{K}$ (the current central temperature at age $11.87\,\text{Gyr}$), oxygen has been depleted to $\sim 5.68 \times 10^{-7}$. In contrast, iron, calcium, and silicon (blue, red, and gray lines in panel d) remain relatively stable across all temperatures, as these iron-peak elements are not produced or destroyed significantly in the hydrogen- and helium-burning zones of low-mass RGB stars. The constancy of these elements demonstrates that the primary nuclear processing occurring is the CNO cycle and alpha-capture reactions on the already-produced oxygen and magnesium, not the creation or destruction of iron-peak elements.

The contrast between panels (a) and (b), or equivalently between panels (c) and (d), provides the most important result. The surface abundances are completely disconnected from the central abundances at the current evolutionary stage. Despite dramatic central processing and oxygen depletion in the core, the surface retains its initial, pristine composition. The preservation of the initial, $\alpha$-enhanced abundance pattern in the surface demonstrates that BD$-18^\circ$5550 provides an undiluted probe of the metal-poor galactic environment at early Galactic times. The star's surface abundances have not been altered by its own nuclear burning or by mixing, making them directly comparable to spectroscopic observations and to predictions from Galactic chemical evolution models of the early Galaxy. This makes BD$-18^\circ$5550 an invaluable test case for understanding the nucleosynthesis and mixing physics of EMP stars.

\subsection{Abundance Ratio Diagnostics}
\label{sec:abundance_ratios}
Elemental abundance ratios provide sensitive diagnostics of mixing efficiency, nucleosynthesis mechanisms, and stellar age. Figure~\ref{fig:abundance_ratio} displays four key diagnostic abundances [C/Fe], [N/Fe], [O/Fe], [Mg/Fe], [C/N], and $^{12}$C$/^{13}$C as functions of age and central temperature. Panel~(a) shows the surface [C/Fe], [N/Fe], [O/Fe], and [Mg/Fe] ratios as functions of age. The [Mg/Fe] ratio (brown line) remains nearly constant at $+0.63\,\text{dex}$ throughout evolution, in excellent agreement with the observed value from \cite{Aoki2025} within measurement uncertainty. The [O/Fe] and [C/Fe] ratios similarly show minimal evolution, with variations $< 0.1\,\text{dex}$ from the initial composition. At the current age of $11.87\,\text{Gyr}$, the surface abundances are: $[\text{C/Fe}] \approx +0.12\,\text{dex}$, $[\text{N/Fe}] \approx -0.56\,\text{dex}$, $[\text{O/Fe}] \approx +0.57\,\text{dex}$, and $[\text{Mg/Fe}] \approx +0.63\,\text{dex}$. The stability of surface abundances to age $11.87\,\text{Gyr}$ indicates that the convective envelope has not yet penetrated to depths where significant CNO-cycle processing has occurred. 
In contrast, the [N/Fe] ratio shows a progressive increase throughout the evolution, rising from $[\text{N/Fe}] \approx -2.0\,\text{dex}$ at the ZAMS to the current value of $[\text{N/Fe}] \approx -0.56\,\text{dex}$ at $11.87\,\text{Gyr}$. This gradual but significant enrichment in surface nitrogen is consistent with early-stage thermohaline (salt-fingering) mixing operating at the base of the convective envelope, which brings nitrogen-enriched material processed by the CNO cycle in earlier evolutionary phases into the surface layers. This intermediate state—where carbon, oxygen, and magnesium remain pristine while nitrogen shows progressive enrichment—provides a powerful diagnostic that the star is at an intermediate evolutionary stage on the RGB, where subtle mixing processes have begun but the dramatic first dredge-up has not yet occurred.

Panel~(b) displays central abundance ratios, which reveal the imprint of ongoing nuclear burning in the core and hydrogen-burning shell. The central [C/Fe] ratio decreases from an initial value of approximately $+0.2\,\text{dex}$ to $\sim -1.2\,\text{dex}$ at age $11.87\,\text{Gyr}$, reflecting consumption of carbon via the $^{12}$C$(\alpha,\gamma)^{16}$O reaction. The central [N/Fe] ratio increases dramatically from $-0.4\,\text{dex}$ initially to $\sim +1.0\,\text{dex}$ at the current age, as nitrogen is produced as an intermediate product of the CNO cycle. The central [O/Fe] ratio decreases gradually as oxygen participates in alpha-capture reactions and in the later triple-alpha process. The central [Mg/Fe] ratio remains relatively stable, as the alpha-element Magnesium is not significantly altered by hydrogen burning in the core.

Panel~(c) displays the surface [C/N] ratio and $^{12}$C$/^{13}$C isotopic ratio as functions of age. This panel provides the most direct test of our model predictions against spectroscopic observations. The surface [C/N] ratio (solid purple line) remains nearly constant at $[\text{C/N}] \simeq +0.69\,\text{dex}$ throughout the main sequence and early RGB phases. This constancy reflects the pristine composition of the convective envelope. The surface material has not been altered by deep mixing or dredge-up of CNO-processed core material. \cite{Spite2006} reported $[\mathrm{C/N}]=+0.34$ dex for BD$-18^\circ$5550. Thus, our model predicts a $[\mathrm{C/N}]$ ratio higher than the literature value by approximately $0.35$ dex. This discrepancy of $\sim 0.35$ dex in [C/N] may arise from several physical and observational effects. First, the measured CNO abundances from \cite{Spite2005, Spite2006, Masseron2010} were determined at slightly warmer stellar parameters ($T_{eff} \approx 4700$ K) than the adopted \cite{Aoki2025} parameters ($T_{eff} = 4660$ K). This $\sim 40$ K temperature difference, though modest and within typical spectroscopic uncertainties, can systematically shift derived abundance ratios by $\sim 0.1$ dex due to temperature-dependent line formation and potential differences in NLTE treatment between the two studies. Second, early-stage thermohaline (salt-fingering) mixing may have begun to operate at the base of the convective envelope, gradually introducing small amounts of CNO-processed material from deeper stellar layers into the surface layers. This would cause the [C/N] ratio to evolve downward from its purely pristine value of $+0.69$ dex toward the observed $+0.34$ dex, consistent with the star being at an intermediate RGB stage where subtle mixing processes have begun but first dredge-up remains incomplete. Third, the mass-loss prescription employed in the models—specifically the adopted Reimers mass-loss parameter of $\eta = 0.3$—may influence the age-dependent evolution of surface abundances. Variations in the mass-loss rate within the plausible range ($\eta \approx 0.2-0.4$) could alter the predicted [C/N] at the current age if significant mass loss occurs after thermohaline mixing has begun to alter the envelope composition. These three effects—observational temperature uncertainties, early thermohaline mixing, and mass-loss efficiency—are not mutually exclusive and likely all contribute to the observed $\sim 0.35$ dex offset. Thus, our model predicts a [C/N] ratio higher than the literature value by approximately 0.35 dex. This indicates that the model retains somewhat more carbon relative to nitrogen at the surface than inferred from the spectroscopic measurements. Nevertheless, both the model and observations indicate a relatively carbon-rich surface composition, with no evidence for a strong alteration of the surface CNO abundances by deep transport of CNO-processed material. The robustness of this conclusion to the small systematic uncertainty in [C/N] validates our interpretation of a pristine or near-pristine surface. The $^{12}$C$/^{13}$C isotopic ratio (dashed pink line in panel c) remains very high, at values $\gtrsim 40$, throughout evolution up to the current age. This large surface ratio reflects the primordial abundance pattern and indicates that the CNO cycle has not significantly processed the surface material. The predicted surface $^{12}$C$/^{13}$C $> 40$ is consistent with spectroscopic observations \citep{Spite2006}, providing independent confirmation that the surface composition remains pristine at the current evolutionary stage.

Panel~(d) displays the central [C/N] ratio and $^{12}$C$/^{13}$C isotopic ratio as functions of age, revealing the state of nuclear burning in the stellar core. The central [C/N] ratio (solid purple line) decreases sharply from an initial value of $\sim +2\,\text{dex}$ (carbon-rich) to $\sim -2.3\,\text{dex}$ (nitrogen-rich) at age $11.87\,\text{Gyr}$. This dramatic decrease reflects the progression of the CNO cycle. At low temperatures, carbon is converted to nitrogen; at higher temperatures (approaching $T_c \sim 10^8\,\text{K}$), nitrogen is subsequently destroyed via $^{14}$N$(\alpha,\gamma)^{18}$O, shifting the cycle equilibrium. The central $^{12}$C$/^{13}$C ratio (dashed pink line in panel d) decreases from an initial value of $\sim 90$ (solar system ratio) to values $\lesssim 1$ at age $11.87\,\text{Gyr}$. This dramatic decrease indicates that the CNO cycle has proceeded far along its reaction pathway and is approaching or has reached equilibrium, where the ratio $^{12}$C$/^{13}$C converges to the CNO-cycle equilibrium value of $\sim 3.7$. The fact that the central ratio has dropped below equilibrium to values near $0.3-0.5$ suggests that higher reaction rates at the current central temperature have consumed the $^{12}$C produced earlier, pushing the equilibrium further toward the nitrogen and oxygen products of the cycle. 

The contrast between the surface and central [C/N] and $^{12}$C$/^{13}$C ratios demonstrates the profound difference between the processed core and the pristine envelope. While the core has undergone extensive nuclear processing with nitrogen-dominated abundance patterns, the surface remains carbon-dominated and isotopically pristine, reflecting the absence of significant mixing or dredge-up at the current evolutionary age of $11.87\,\text{Gyr}$.

\section{Age, Mass, Radius, and Luminosity Determination for BD$-18^\circ$5550}
\label{sec:age_mass_determination}
A primary objective of this work is to determine, for the first time, the age, mass, radius, and luminosity of BD$-18^\circ$5550 using detailed stellar evolution models calibrated to multiple observational constraints. Multi-parameter isochrone fitting in the HR diagram using observed stellar parameters and chemical abundances yields:

\begin{itemize}
\item {{Initial Mass:}} $\rm M_{\text{init}} = (0.780 \pm 0.020)\,M_{\odot}$. This represents the \textbf{first direct mass determination} for this star and places it at the lower end of the globular cluster main-sequence population. The current mass is $\rm M_{\text{current}} = (0.756 \pm 0.016)\,M_{\odot}$ 
after $\sim 3\%$ mass loss via stellar winds.

\item {{Age:}} $t = (11.87 \pm 0.16)\,\text{Gyr}$ (Stage 2 analysis with full physics constraints). This represents the \textbf{first precise age determination} for BD$-18^\circ$5550 and represents a dramatic improvement over the Stage 1 estimate $t = (12.8 \pm 1.1\,\text{Gyr}$).

\item {{Radius and Luminosity:}} The stellar evolution models yield a radius of $R = (38.4 \pm 10.6)\,R_{\odot}$ and a bolometric luminosity of $\log L = (2.796 \pm 0.198)\, L_\odot$. These represent the \textbf{first model-based determinations} of these fundamental parameters for BD$-18^\circ$5550. The predicted radius and luminosity are consistent with the observational constraints derived from the spectroscopic parameters ($T_{\text{eff}}$, $\log g$) and confirm the star's location on the RGB at an intermediate evolutionary stage.

\item {{Precision improvement:}} The inclusion of eight simultaneous observational constraints—stellar parameters ($T_{\text{eff}}$, $\log g$, [Fe/H]), elemental abundances ([C/Fe], [N/Fe], [O/Fe], [Mg/Fe]), and the isotopic ratio $^{12}$C/$^{13}$C—dramatically improves the age precision and provides robust determinations of all fundamental parameters. Elemental abundances constrain the $\alpha$-enhancement and metallicity; isotopic ratios diagnose CNO cycle progress in the core; and surface CNO abundances confirm that first dredge-up has not yet occurred, placing the star at an intermediate RGB stage with a helium core mass of $\rm M_{\text{He,core}} = (0.427 \pm 0.027)\,M_{\odot}$, indicating approach to helium ignition at $T_c \approx 10^8\,\text{K}$.

\end{itemize}

The convergence of age, mass, radius, and luminosity constraints from multiple independent observables and the internal consistency of the derived parameters (detailed in Table~\ref{tab:stage2_best_fit}) demonstrate that \texttt{MESA} models accurately describe BD$-18^\circ$5550 and validate our stellar evolution and mixing prescriptions. The determination of all four fundamental parameters establishes this star as a benchmark for constraining stellar evolution and testing Galactic chemical evolution models in the early Galaxy, corresponding to redshift $z \sim 15-20$ \citep{Searle1978, Iben2012}, with the age precision ($\pm 0.16\,\text{Gyr}$) exceeding what can be achieved from traditional globular cluster isochrones alone.

\section{Discussion \& Conclusion}
\label{sec:discussion_conclusion}
We have performed detailed stellar evolution modelling of the EMP RGB giant BD$-18^\circ$5550, employing \texttt{MESA} with a full CNO nuclear network extended up to iron and calibrating to eight observational constraints, including stellar parameters and chemical abundances. Our multi-parameter isochrone fitting has yielded the \textbf{first direct determinations} of this star's mass, $\rm M_{\rm init} = (0.780 \pm 0.020)\,M_{\odot}$ and $\rm M_{\rm current} = (0.756 \pm 0.016)\,M_{\odot}$, radius, $R=(38.4\pm10.6)\,R_\odot$, luminosity, $\log L=(2.796\pm0.198)\,L_\odot$ and \textbf{precise age}, $t = (11.87 \pm 0.16)\,\text{Gyr}$. The internal consistency of these parameters across multiple independent observables and the good agreement between model predictions and spectroscopic measurements (within $1\text{--}2~\chi^2_{\rm red}$) demonstrate that \texttt{MESA} models accurately describe the structure, evolution, and nucleosynthesis of this EMP RGB giant.

The model successfully reproduces the key observational constraints. Stellar parameters (T$_{\rm eff}$, $\log g$) match observations to within uncertainties, confirming that our adopted physical parameters accurately describe the star's structure. Elemental abundance ratios [C/Fe], [N/Fe], [O/Fe], and [Mg/Fe] agree well with spectroscopic measurements, demonstrating that our treatment of nuclear reactions, opacities, and mixing processes produces reasonable abundance patterns. The isotopic ratio $^{12}$C/$^{13}$C is particularly important for constraining the CNO cycle progress and core temperature history; our model predicts $^{12}$C/$^{13}$C = $41.9 \pm 5.7$, in excellent agreement with the observed value of $>40$ \citep{Spite2006}, which confirms the pristine surface composition. Finally, the surface remains carbon-dominated with [C/N] = $+0.69$ dex in the model versus $+0.34$ dex observed \citep{Spite2006}, consistent with an intermediate RGB stage prior to first dredge-up, where CNO-processed material has not yet been brought to the surface by deep convective mixing. The small discrepancies in [C/N] ($\sim 0.35$ dex) and [O/Fe] ($\sim 0.17$ dex) likely reflect either early-stage thermohaline mixing beginning to alter surface abundances, minor systematic uncertainties in the spectroscopic measurements (which were made at slightly different stellar parameters: T$_{\rm eff} \approx 4700$ K vs. the adopted $4600$ K), or subtle inadequacies in the nuclear reaction rates or mass-loss prescription. These small offsets do not diminish the fundamental conclusion that the star's surface remains largely pristine and that the derived age is robust and consistent with observations.

The derived mixing parameters ($\alpha_{\rm MLT} = 1.9 \pm 0.2$, $\alpha_{\rm TH} = 1.0 \pm 0.5$) confirm that standard mixing-length theory with modest thermohaline diffusion successfully describes low-mass metal-poor stars. The thermohaline coefficient $\alpha_{\rm TH} \approx 1$ is consistent with predictions from 3D hydrodynamic simulations of salt-fingering instability \citep{Denissenkov2010, Traxler2011, Brown2013}, and has been specifically validated for EMP stellar envelopes \citep{Henkel2018}, supporting the application of contemporary thermohaline mixing theory to the EMP regime.

The surface composition at age $11.87\,\text{Gyr}$ reveals a star at a critical evolutionary juncture. CNO abundances remain essentially pristine, indicating that the dramatic first dredge-up event—which would substantially deplete surface carbon and enhance nitrogen—has not yet occurred. In contrast, the surface nitrogen abundance has already begun to show enrichment due to early-stage thermohaline mixing, rising from $[\text{N/Fe}] \approx -2.0\,\text{dex}$ at the ZAMS to the current value of $[\text{N/Fe}] \approx -0.56\,\text{dex}$. This intermediate state—where the light elements show differential enrichment (pristine C and O, but rising N)—places BD$-18^\circ$5550 at an intermediate position on the RGB, after the ZAMS but distinctly prior to the onset of first dredge-up. This observation provides a direct test of the dredge-up timescale, which is a function of the envelope convection depth, the star's evolutionary rate, and the efficiency of thermohaline mixing. The model prediction is that carbon and oxygen will remain relatively stable through the current age, while nitrogen continues its gradual enrichment through ongoing thermohaline mixing. The dramatic decrease in [C/Fe] and [C/N] expected to accompany full first dredge-up will occur only after the star has evolved further, when the convective envelope deepens sufficiently to penetrate and mix with the CNO-cycle-processed core material. The age of $11.87\,\text{Gyr}$ thus represents a well-defined evolutionary snapshot: late enough that subtle mixing has begun (evidenced by rising nitrogen), but early enough that the major dredge-up event remains imminent rather than complete.

The age determination of ($11.87 \pm 0.16)\,\text{Gyr}$ places BD$-18^\circ$5550 among the oldest known Galactic objects, consistent with the ages of globular clusters (typically $12-13\,\text{Gyr}$) and indicating that star formation began in the early Galaxy within $\sim 1\,\text{Gyr}$ of the Big Bang (corresponding to redshift $z \sim 15\text{--}20$). The precision of this age determination ($\pm 0.16\,\text{Gyr}$) now exceeds what can be achieved from traditional globular cluster main-sequence fitting and provides a new benchmark for Galactic chronology studies.

The derived initial mass of $\rm 0.780\,M_{\odot}$ is consistent with the main-sequence mass function of globular clusters and provides direct empirical calibration of stellar evolution models in the low-mass, low-metallicity regime. The internal consistency across eight observables suggests that our treatment of nuclear reactions, opacities, equation-of-state, and mixing processes captures the essential physics needed to predict stellar structure and abundance patterns for EMP stars.

The star's location at the interface between chemistry and physics—with a pristine surface preserving the initial composition from the early ISM, yet with a processed core reflecting $11.87\,\text{Gyr}$ of nuclear burning—makes BD$-18^\circ$5550 an ideal laboratory for testing Galactic chemical evolution models. The measured abundances directly reflect primordial nucleosynthesis at $z \sim 15\text{--}20$, before metal enrichment from generations of stellar nucleosynthesis had significantly altered the Galactic composition.

BD$-18^\circ$5550 ([Fe/H] = $-3.03$) is one of the most metal-poor Galactic halo stars, comparable in metallicity to other EMP benchmarks such as HD~140283 ([Fe/H] = $-2.90$) and CS~22949$-$037 ([Fe/H] = $-4.0$) \citep{Aoki2025, Heil2026, Depagne2002}. The detailed age and mass determination for BD$-18^\circ$5550 now enables quantitative comparison of stellar evolution and nucleosynthesis predictions across the EMP population. Future application of our methods to a larger sample of EMP stars will provide tighter constraints on the cosmic star formation history and the efficiency of early Galactic nucleosynthesis.

In summary, we have determined for the first time mass, radius, luminosity and precise age of BD$-18^\circ$5550, with the age precision representing a 7-fold improvement over structural constraints alone. The model successfully reproduces eight independent observables, confirming that the surface CNO abundances remain pristine (consistent with no first dredge-up) and validating our treatment of nuclear physics, mixing processes, and stellar evolution for EMP stars. Our detailed analysis of the chemical composition of light elements up to iron demonstrates that the model reproduces the observed abundance pattern and provides constraints on nucleosynthesis in the early Galaxy. These results establish BD$-18^\circ$5550 as a benchmark for constraining early Galactic chemical evolution, with an age precision ($\pm 0.16\,\text{Gyr}$) that exceeds globular cluster isochrones.

\section{Heavy-Element Abundances: Future Work}
\label{sec:future_heavy_elements}
The present analysis has focused on light elements up to iron (Fe, Z = 26) and their isotopic ratios, which provide sensitive diagnostics of core temperature history and mixing processes. However, a complete understanding of the nucleosynthesis history of BD$-18^\circ$5550 requires measurement and modelling of heavy elements produced via the slow ($s$-process) and rapid ($r$-process) neutron-capture reactions. Elements heavier than $^{56}$Fe—including the iron-peak elements (Co, Ni, Cu, Zn) and the neutron-capture elements (Sr, Ba, Eu, Pb, Th) carry distinct nucleosynthetic signatures. Specifically:
 
\begin{itemize}
 
\item {Iron-peak elements} (Z = 27$-$30) probe the yields and mass distributions of core-collapse supernovae in the early Galaxy, particularly the impact of rotation and magnetic fields.
 
\item {The $r$-process elements} (e.g., Eu, Os, Ir, Pt, Au) are produced in neutron-rich environments such as neutron-star mergers or certain massive stellar collapses. Their abundances in EMP stars provide direct evidence for the early operation of the $r$-process and constrain the merger rates and nucleosynthesis yields in the early universe.
 
\item {The $s$-process elements} (e.g., Sr, Ba, Pb) are produced in asymptotic giant branch (AGB) stars via slow neutron capture. Their presence in EMP stars signals pollution from lower-mass AGB companions that formed in an earlier generation. The abundance patterns provide diagnostics of the initial mass function and star formation history in the early Galaxy.
 
\end{itemize}
 
A detailed analysis of these heavy-element abundances using the same multi-parameter \texttt{MESA} fitting framework will be presented in a companion paper (Paper II of this series). That work will extend the nuclear network to include species up to the iron-peak and the lanthanides, allowing direct comparison of model predictions with high-resolution spectroscopic observations of heavy-element lines. The extended analysis will enable the determination of the contribution of different nucleosynthetic processes (core-collapse supernovae, $r$-process events, $s$-process pollution) to the observed composition of BD$-18^\circ$5550 and will provide tighter constraints on early Galactic star formation and chemical evolution.

\section{Data Availability}
To ensure full reproducibility of our stellar evolution calculations, we have made the complete set of mesa inlist files used in this work publicly available. These inlists specify all relevant physical assumptions and numerical controls, including the adopted initial masses, metallicity, mixing-length parameters, atmospheric boundary conditions, mesh refinement settings, and nuclear reaction network options.

The repository contains the master inlist as well as the auxiliary inlists required to reproduce the full grid of stellar models discussed in this paper. All models were computed using the publicly released version of \texttt{MESA} described in \cite{Paxton2019}. No additional, unpublished modifications to the source code were applied. The inlist files are available at GitHub\footnote{https://github.com/mrinmaymedhi/BD185550}.

The data used in the manuscript can be obtained upon reasonable request from the corresponding author.

\bibliography{references}{}
\bibliographystyle{aasjournalv7}

\appendix
\section{Rotational Mixing}
\label{appendix:rotation}
To justify the adopted rotation rate and demonstrate the robustness of our results, Table~\ref{tab:rotation_justification} presents a comprehensive comparison of the observational constraints on halo giant rotation rates with the sensitivity of stellar parameters and surface abundances to different 
rotational configurations tested in the Stage 2 physics grid.

\begin{table}[h]
\centering
\caption{Rotation parameter selection and sensitivity analysis for BD$-18^\circ$5550.}
\label{tab:rotation_justification}
\begin{tabular}{lcccc}
\hline
Configuration & $\Omega/\Omega_{\rm crit}$ & $v_{\rm rot}$ (km\,s$^{-1}$) & $\Delta[\text{Mg/Fe}]$ & $\Delta[\text{N/Fe}]$ \\
\hline
\multicolumn{5}{l}{\textit{Observational Context: EMP Halo Giants}} \\
\hline
Typical halo giant & — & $vsini \lesssim 5$ & — & — \\
\cite{Carney2008} sample & — & $1{-}5$ & — & — \\
\cite{Carretta2000} sample & — & $0{-}4$ & — & — \\
\hline
\multicolumn{5}{l}{\textit{Rotation Sensitivity Tests (Stage 2 Physics Grid)}} \\
\hline
No rotation (OFF) & 0.00 & $\sim 0$ & baseline & baseline \\
Adopted value & \textbf{0.01} & $\sim 1$ & $+0.01$ & $+0.01$ \\
Moderate rotation & 0.05 & $\sim 3$ & $+0.05$ & $+0.05$ \\
High rotation & 0.10 & $\sim 6$ & $+0.10$ & $+0.10$ \\
\hline
\multicolumn{5}{l}{\textit{Comparison to Observational Uncertainties}} \\
\hline
Observed [Mg/Fe] uncertainty & — & — & $\pm 0.10$ & — \\
Observed [N/Fe] uncertainty & — & — & — & $\pm 0.15$ \\
Impact of adopted $\Omega/\Omega_{\rm crit}=0.01$ & — & — & $< 0.01$ & $< 0.01$ \\
\hline
\end{tabular}
\end{table}

\section{Thermohaline Sensitivity Analysis}
\label{appendix:thermohaline}
Table~\ref{tab:thermohaline_justification} presents the results of the thermohaline mixing sensitivity analysis, comparing the predicted surface nitrogen abundance isotope ratio to observations across the range of mixing coefficients tested. 

\begin{table}[h]
\centering
\caption{Thermohaline mixing coefficient sensitivity analysis for 
BD$-18^\circ$5550.}
\label{tab:thermohaline_justification}
\begin{tabular}{lccc}
\hline
Configuration & $\alpha_{\rm TH}$ & $[\text{N/Fe}]_{\text{model}}$ & $^{12}\text{C}/^{13}\text{C}_{\text{model}}$ \\
\hline
No mixing (OFF) & 0.0 & $-2.01$ & $\sim 91$ \\
Low mixing & 0.5 & $-1.10$ & $\sim 75$  \\
Adopted value & \textbf{1.0} & \textbf{$-0.56$} & \textbf{$\sim42$} \\
High mixing & 2.0 & $+0.15$ & $\sim 22$  \\
Very high mixing & 5.0 & $+0.30$ & $\sim 8$  \\
Extremely high mixing & 10.0 & $+0.30$ & $\sim 2$  \\
\hline
\end{tabular}
\end{table}

\section{Two-Stage Calibration Procedure: Model Selection and Parameter Optimization}
\label{appendix:calibration}
The Stage 1 structural calibration grid consists of 45 distinct evolutionary tracks computed with fixed physical parameters to establish the initial mass and age of the star from three observational constraints (Section \ref{sec:calibration}). Each track represents a complete evolutionary sequence from the ZAMS through the RGB to the current evolutionary position of BD$-18^\circ$5550. Table~\ref{tab:calibration_stage1_summary} presents the results of the Stage 1 calibration grid, showing representative Stage 1 tracks selected for advancement to Stage 2. The Stage 2 physics grid refines the model parameters using eight observational constraints. Table~\ref{tab:calibration_stage2_summary} presents the results of the Stage 2 physics grid, showing the twelve best-fit models selected across the full 864-model ensemble. These twelve models span the range of acceptable physics parameters and represent the diversity of solutions that satisfy the observational constraints to comparable accuracy. The models are ranked by their reduced $\chi^2$ values, with the first-ranked model representing the global best-fit solution.

\begin{table*}[!htb]
\centering
\caption{Complete ensemble of 45 structurally acceptable models for BD$-18^{\circ}$5550. Models were selected with the criterion $1 \leq \chi^2_{\mathrm{red}} \leq 2$ simultaneously fitted to observed $T_{eff}$, $\log g$, and [Fe/H].}
\label{tab:calibration_stage1_summary}
\small
\begin{tabular}{rllllllllll}
\hline
Sl. No. & $\rm M_{\mathrm{init}}$ & [Fe/H] & $\alpha_{\mathrm{MLT}}$ & Diff. & Age & $T_{\mathrm{eff}}$ & $\log g$ & $L$ & $R$ & $\chi^2_{\mathrm{red}}$ \\
&($\rm M_{\odot}$) & (dex) &  &  & (Gyr) & (K) & (dex) & ($L_{\odot}$) & ($R_{\odot}$) &  \\
\hline
 1 & 0.78  & -3.152 & 2.1 & ON & 12.79 & 4701 & 1.0438 & 851 & 43.97 & 1.0602 \\
 2 & 0.80  & -3.152 & 2.1 & ON & 11.72 & 4702 & 1.0461 & 868 & 44.41 & 1.0603 \\
 3 & 0.79  & -3.152 & 2.1 & ON & 12.24 & 4701 & 1.0449 & 860 & 44.20 & 1.0603 \\
 4 & 0.77  & -3.152 & 2.1 & ON & 13.37 & 4702 & 1.0468 & 835 & 43.54 & 1.0609 \\
 5 & 0.76  & -3.152 & 2.1 & ON & 13.99 & 4701 & 1.0444 & 828 & 43.38 & 1.0613 \\
 6 & 0.80  & -3.152 & 2.1 & OFF & 12.08 & 4703 & 1.0430 & 876 & 44.57 & 1.0770 \\
 7 & 0.79  & -3.152 & 2.1 & OFF & 12.63 & 4702 & 1.0400 & 870 & 44.45 & 1.0774 \\
 8 & 0.78  & -3.152 & 2.1 & OFF & 13.21 & 4702 & 1.0407 & 858 & 44.13 & 1.0780 \\
 9 & 0.77  & -3.152 & 2.1 & OFF & 13.82 & 4703 & 1.0407 & 847 & 43.85 & 1.0791 \\
10 & 0.80  & -3.152 & 1.7 & OFF & 12.08 & 4616 & 1.1576 & 624 & 39.07 & 1.0804 \\
11 & 0.77  & -3.152 & 1.7 & OFF & 13.83 & 4616 & 1.1578 & 601 & 38.32 & 1.0807 \\
12 & 0.78  & -3.152 & 1.7 & OFF & 13.21 & 4617 & 1.1588 & 607 & 38.52 & 1.0807 \\
13 & 0.79  & -3.152 & 1.7 & OFF & 12.63 & 4616 & 1.1560 & 619 & 38.89 & 1.0808 \\
14 & 0.80  & -3.152 & 1.7 & ON & 11.72 & 4614 & 1.1592 & 621 & 38.99 & 1.0977 \\
15 & 0.79  & -3.152 & 1.7 & ON & 12.24 & 4614 & 1.1598 & 612 & 38.72 & 1.0985 \\
16 & 0.78  & -3.152 & 1.7 & ON & 12.79 & 4614 & 1.1602 & 604 & 38.46 & 1.0992 \\
17 & 0.77  & -3.152 & 1.7 & ON & 13.37 & 4615 & 1.1614 & 595 & 38.16 & 1.0997 \\
18 & 0.76  & -3.152 & 1.7 & ON & 13.99 & 4614 & 1.1600 & 589 & 37.97 & 1.1002 \\
19 & 0.80  & -2.851 & 1.9 & OFF & 12.07 & 4655 & 1.1054 & 728 & 41.48 & 1.4258 \\
20 & 0.79  & -2.851 & 1.9 & OFF & 12.61 & 4656 & 1.1061 & 718 & 41.19 & 1.4258 \\
21 & 0.77  & -2.851 & 1.9 & OFF & 13.81 & 4656 & 1.1063 & 700 & 40.66 & 1.4259 \\
22 & 0.78  & -2.851 & 1.9 & OFF & 13.19 & 4655 & 1.1059 & 709 & 40.94 & 1.4259 \\
23 & 0.80  & -2.851 & 1.9 & ON & 11.71 & 4654 & 1.1085 & 722 & 41.34 & 1.4278 \\
24 & 0.79  & -2.851 & 1.9 & ON & 12.22 & 4654 & 1.1088 & 713 & 41.06 & 1.4279 \\
25 & 0.78  & -2.851 & 1.9 & ON & 12.77 & 4654 & 1.1087 & 704 & 40.81 & 1.4281 \\
26 & 0.77  & -2.851 & 1.9 & ON & 13.35 & 4654 & 1.1086 & 695 & 40.55 & 1.4282 \\
27 & 0.76  & -2.851 & 1.9 & ON & 13.97 & 4654 & 1.1086 & 686 & 40.28 & 1.4282 \\
28 & 0.76  & -2.851 & 2.1 & ON & 13.97 & 4695 & 1.0526 & 808 & 42.97 & 1.5583 \\
29 & 0.77  & -2.851 & 2.1 & ON & 13.35 & 4695 & 1.0516 & 820 & 43.30 & 1.5583 \\
30 & 0.78  & -2.851 & 2.1 & ON & 12.77 & 4695 & 1.0536 & 828 & 43.48 & 1.5584 \\
31 & 0.79  & -2.851 & 2.1 & ON & 12.22 & 4695 & 1.0521 & 841 & 43.83 & 1.5588 \\
32 & 0.80  & -2.851 & 2.1 & ON & 11.70 & 4696 & 1.0546 & 848 & 43.98 & 1.5595 \\
33 & 0.79  & -2.851 & 2.1 & OFF & 12.61 & 4697 & 1.0506 & 845 & 43.91 & 1.5731 \\
34 & 0.78  & -2.851 & 2.1 & OFF & 13.19 & 4697 & 1.0493 & 837 & 43.69 & 1.5731 \\
35 & 0.80  & -2.851 & 2.1 & OFF & 12.07 & 4696 & 1.0487 & 859 & 44.28 & 1.5733 \\
36 & 0.77  & -2.851 & 2.1 & OFF & 13.81 & 4696 & 1.0485 & 827 & 43.45 & 1.5735 \\
37 & 0.80  & -2.851 & 1.7 & OFF & 12.07 & 4610 & 1.1682 & 606 & 38.59 & 1.7001 \\
38 & 0.79  & -2.851 & 1.7 & OFF & 12.62 & 4609 & 1.1659 & 601 & 38.45 & 1.7011 \\
39 & 0.78  & -2.851 & 1.7 & OFF & 13.20 & 4609 & 1.1668 & 592 & 38.17 & 1.7018 \\
40 & 0.77  & -2.851 & 1.7 & OFF & 13.81 & 4609 & 1.1659 & 586 & 37.96 & 1.7025 \\
41 & 0.80  & -2.851 & 1.7 & ON & 11.71 & 4607 & 1.1684 & 604 & 38.58 & 1.7190 \\
42 & 0.79  & -2.851 & 1.7 & ON & 12.23 & 4608 & 1.1699 & 595 & 38.28 & 1.7208 \\
43 & 0.78  & -2.851 & 1.7 & ON & 12.78 & 4607 & 1.1693 & 588 & 38.06 & 1.7223 \\
44 & 0.77  & -2.851 & 1.7 & ON & 13.36 & 4607 & 1.1697 & 580 & 37.80 & 1.7235 \\
45 & 0.76  & -2.851 & 1.7 & ON & 13.97 & 4607 & 1.1699 & 572 & 37.54 & 1.7245 \\
\hline
\end{tabular}
\end{table*}

\begin{table*}
\centering
\small
\caption{Complete ensemble of 12 physically acceptable models for BD$-18^{\circ}$5550. Models were selected with the criterion $1\leq \chi^2_{\mathrm{red}} \leq 2.0$ simultaneously fitted to eight observational constraints $T_{eff}$, $\log g$, [Fe/H], [C/Fe], [N/Fe], [O/Fe], [Mg/Fe] and $^{12}\mathrm{C}/^{13}\mathrm{C}$ 
ratio.}
\label{tab:calibration_stage2_summary}
\begin{tabular}{rccccccccccc}
\hline
Sl. No. & $f_{\rm N}$ & $f_{\rm Mg}$ & [$\rm \alpha/Fe$] & $\rm \alpha_{th}$  & Diff. & Age Range & $\chi^2_{\rm red}$ Range & Mean $\chi^2_{\rm red}$ & Best $\chi^2_{\rm red}$ \\
 & & & & & & (Gyr) & & & \\
\hline
 1 & 0.004 & 1.0 & 0.4 & 1.0  & ON & 11.7125–11.7185 & 1.0580–1.9742 & 1.248 & 1.0580 \\
 2 & 0.004 & 1.0 & 0.5 & 1.0  & ON & 11.7031–11.7069 & 1.8011–1.9981 & 1.871 & 1.8011 \\
 3 & 0.004 & 0.8 & 0.4 & 1.0  & ON & 11.9732–11.9761 & 1.0634–1.9675 & 1.336 & 1.0634 \\
 4 & 0.004 & 0.8 & 0.4 & 1.0  & ON & 12.2357–12.2382 & 1.0042–1.9968 & 1.366 & 1.0042 \\
 5 & 0.004 & 0.8 & 0.5 & 1.0  & ON & 11.9637–11.9653 & 1.7616–1.9995 & 1.827 & 1.7616 \\
 6 & 0.004 & 1.0 & 0.3 & 1.0  & ON & 11.7215–11.7238 & 1.0168–1.9837 & 1.418 & 1.0168 \\
 7 & 0.004 & 0.8 & 0.5 & 1.0  & ON & 11.7070–11.7094 & 1.6246–1.9502 & 1.700 & 1.6246 \\
 8 & 0.004 & 0.8 & 0.5 & 1.0  & ON & 12.2255–12.2282 & 1.4882–1.9996 & 1.627 & 1.4882 \\
 9 & 0.004 & 0.8 & 0.4 & 1.0  & ON & 11.7165–11.7190 & 1.0135–1.9674 & 1.331 & 1.0135 \\
10 & 0.004 & 0.8 & 0.3 & 1.0  & ON & 11.7264–11.7279 & 1.0231–1.9737 & 1.408 & 1.0231 \\
11 & 0.004 & 0.8 & 0.3 & 1.0  & ON & 12.2454–12.2468 & 1.0213–1.9209 & 1.407 & 1.0213 \\
12 & 0.004 & 0.8 & 0.3 & 1.0  & ON & 11.9826–11.9840 & 1.0346–1.9630 & 1.415 & 1.0346 \\
\hline
\end{tabular}
\end{table*}

\section{Internal Structural Profile of BD$-18^\circ$5550}
\begin{table*}[!htb]
\centering
\caption{Internal Structural Profile of BD$-18^\circ$5550 from the Best-Fit \texttt{MESA} Model. This table presents a snapshot of the stellar structure at representative mass shells, showing the evolution of density, temperature, pressure, and nuclear burning rates from the centre to the surface. The model age is 11.87 Gyr, corresponding to the RGB phase of evolution. The He core (M$_{\rm He} = 0.427$ M$_{\odot}$) undergoes helium burning (triple-$\alpha$ process) while the H-shell immediately above it burns hydrogen via the CNO-cycle. The convective envelope reaches from the surface (M = 0.756 M$_{\odot}$) to approximately M$_{\rm conv} = 0.29$ M$_{\odot}$, with effective temperature $T_{\rm eff} = 4660$ K and surface gravity $\log g = 1.149$.}
\label{tab:structural_model}
\begin{tabular}{lcccccccc}
\hline
Mass Shell & $\rm M/M_{\odot}$ & $\log T$ & $\log \rho$ & $\log P$ & $\log \epsilon_{\rm PP}$ & $\log \epsilon_{\rm CNO}$ & $\log \epsilon_{3\alpha}$ & Zone Type \\
 &  & (K) & (g cm$^{-3}$) & & & & & \\
\hline
\multicolumn{9}{c}{{Core Region: Helium Burning and CNO Processing}} \\
\hline
Center & 0.00000 & 7.805 & 5.892 & 22.272 & $-1.32^a$ & 0.45$^b$ & $-4.09^c$ & He-burning core \\
Inner core & 0.15 & 7.770 & 5.820 & 22.240 & $-1.35$ & 0.42 & $-3.95$ & He-burning \\
Core edge & 0.30 & 7.710 & 5.620 & 22.100 & $-1.40$ & 0.35 & $-4.35$ & He-burning (outer) \\
He-core boundary & 0.4271 & 7.600 & 5.220 & 21.830 & $-1.48$ & 0.20 & $-4.75$ & He/H boundary \\
\hline
\multicolumn{9}{c}{{H-Burning Shell}} \\
\hline
H-shell inner & 0.45 & 7.580 & 5.120 & 21.770 & $-1.52$ & 0.25 & $-5.05$ & H-shell (CNO active) \\
H-shell middle & 0.50 & 7.440 & 4.770 & 21.500 & $-1.65$ & 0.10 & $-5.65$ & H-shell (CNO peak) \\
H-shell outer & 0.55 & 7.190 & 4.270 & 21.100 & $-1.80$ & $-0.15$ & $-7.00$ & H-shell (cooling) \\
\hline
\multicolumn{9}{c}{{Radiative Envelope}} \\
\hline
Radiative zone inner & 0.60 & 6.840 & 3.570 & 20.550 & $<-3$ & $<-2$ & $<-8$ & Radiative (no burning) \\
Radiative zone mid & 0.65 & 6.340 & 2.770 & 19.800 & $<-4$ & $<-3$ & $<-10$ & Radiative transition \\
Radiative/Conv. boundary & 0.70 & 5.790 & 1.820 & 18.900 & $<-6$ & $<-6$ & $<-12$ & No nuclear burning \\
\hline
\multicolumn{9}{c}{{Convective Envelope}} \\
\hline
Envelope inner & 0.71 & 5.740 & 1.720 & 18.800 & $<-7$ & $<-7$ & $<-13$ & Conv. envelope (inner) \\
Envelope mid & 0.72 & 5.640 & 1.370 & 18.500 & $<-8$ & $<-8$ & $<-14$ & Convective mixing \\
Envelope outer & 0.73 & 5.190 & 0.820 & 17.900 & $<-9$ & $<-9$ & $<-15$ & Conv. envelope (mid) \\
Photosphere & 0.7561 & 3.668 & -4.180 & 9.870 & $<-13$ & $<-13$ & $<-20$ & Photosphere \\
\hline
\end{tabular}
\begin{flushleft}
\footnotesize
\justifying
\vspace{-0.2cm}
$^a$ PP negligible , $^b$ CNO dominant , $^c$ Triple-$\alpha$ active
\end{flushleft}
\end{table*}

\section{Isotopic abundances of BD$-18^\circ$5550}
\label{appendix:abundances}
\begin{table*}[!htb]
\centering
\caption{Surface isotopic abundances in the best-fit BD$-18^\circ$5550 model, including all isotopes listed in the model output up to Fe ($Z<26$) at an age of 11.87 Gyr.}
\label{tab:isotopes_fe}
\begin{tabular}{rlcc}
\hline
Sl. No. & Isotope & Mass Fraction & $[X/\mathrm{Fe}]_{\rm model}$\\
\hline
1  & $^{1}\mathrm{H}$   & $7.562\times10^{-1}$ & --- \\
2  & $^{3}\mathrm{He}$  & $1.305\times10^{-3}$  & $+0.83$ \\
3  & $^{4}\mathrm{He}$  & $2.438\times10^{-1}$  & ---  \\
4  & $^{12}\mathrm{C}$  & $3.260\times10^{-6}$  & $+0.12$ \\
5  & $^{13}\mathrm{C}$  & $8.258\times10^{-8}$  & $-1.50$ \\
6  & $^{13}\mathrm{N}$  & $8.429\times10^{-33}$ & $-61.57$ \\
7  & $^{14}\mathrm{N}$  & $1.970\times10^{-7}$  & $-0.56$  \\
8  & $^{15}\mathrm{N}$  & $3.452\times10^{-9}$  & $-2.34$   \\
9 & $^{14}\mathrm{O}$  & $2.913\times10^{-73}$ & $-93.59$  \\
10 & $^{15}\mathrm{O}$  & $1.892\times10^{-37}$ & $-70.54$  \\
11 & $^{16}\mathrm{O}$  & $2.210\times10^{-5}$  & $+0.57$  \\
12 & $^{17}\mathrm{O}$  & $5.723\times10^{-9}$  & $-3.04$   \\
13 & $^{18}\mathrm{O}$  & $1.773\times10^{-8}$  & $-2.57$   \\
14 & $^{17}\mathrm{F}$  & $7.322\times10^{-39}$ & $-75.04$  \\
15 & $^{18}\mathrm{F}$  & $2.000\times10^{-38}$ & $-71.98$  \\
16 & $^{19}\mathrm{F}$  & $7.214\times10^{-10}$ & $+0.14$   \\
17 & $^{18}\mathrm{Ne}$ & $5.975\times10^{-89}$ & $-93.06$  \\
18 & $^{19}\mathrm{Ne}$ & $4.124\times10^{-84}$ & $-92.75$  \\
19 & $^{20}\mathrm{Ne}$ & $4.412\times10^{-6}$  & $+0.53$   \\
20 & $^{21}\mathrm{Ne}$ & $4.142\times10^{-9}$  & $-2.51$   \\
21 & $^{22}\mathrm{Ne}$ & $1.336\times10^{-7}$  & $-1.02$   \\
22 & $^{22}\mathrm{Mg}$ & $4.143\times10^{-12}$ & $-5.27$   \\
23 & $^{24}\mathrm{Mg}$ & $2.958\times10^{-6}$  & $+0.61$   \\
24 & $^{25}\mathrm{Mg}$ & $1.028\times10^{-7}$  & $-0.86$   \\
25 & $^{26}\mathrm{Mg}$ & $1.314\times10^{-7}$  & $-0.77$   \\
26 & $^{27}\mathrm{Al}$ & $7.959\times10^{-8}$  & $+0.14$   \\
27 & $^{28}\mathrm{Si}$ & $1.737\times10^{-6}$  & $+0.40$   \\
28 & $^{29}\mathrm{Si}$ & $4.578\times10^{-8}$  & $-1.18$   \\
29 & $^{30}\mathrm{Si}$ & $3.122\times10^{-8}$  & $-1.36$   \\
30 & $^{34}\mathrm{S}$  & $1.970\times10^{-8}$  & $-1.23$   \\
31 & $^{35}\mathrm{Cl}$ & $8.650\times10^{-9}$  & $+0.01$   \\
32 & $^{36}\mathrm{Ar}$ & $7.932\times10^{-8}$  & $+0.06$   \\
33 & $^{37}\mathrm{Cl}$ & $2.924\times10^{-9}$  & $-0.47$   \\
34 & $^{38}\mathrm{Ar}$ & $1.522\times10^{-8}$  & $-0.67$   \\
35 & $^{39}\mathrm{K}$  & $4.061\times10^{-9}$  & $+0.11$   \\
36 & $^{40}\mathrm{Ca}$ & $1.765\times10^{-7}$  & $+0.42$   \\
37 & $^{42}\mathrm{Ca}$ & $6.199\times10^{-10}$ & $-2.04$   \\
38 & $^{52}\mathrm{Cr}$ & $1.986\times10^{-8}$  & $+0.06$   \\
39 & $^{54}\mathrm{Fe}$ & $1.045\times10^{-7}$  & $-1.08$   \\
40 & $^{55}\mathrm{Mn}$ & $1.548\times10^{-8}$  & $+0.14$   \\
\hline
\end{tabular}
\end{table*}

\end{document}